\documentclass[a4paper,fleqn,usenatbib]{mnras}

\usepackage{newtxtext,newtxmath}

\usepackage[T1]{fontenc}
\usepackage{ae,aecompl}

\usepackage{graphicx}	
\usepackage{amsmath}	
\usepackage{amssymb}	
\usepackage{mathrsfs}   
\usepackage{xcolor}

\newcommand{\mpch}{\,$h^{-1}$\,Mpc}
\newcommand{\hmpc}{\,$h$\,Mpc$^{-1}$}

\newcommand{\deh}{\delta_\mathrm{h}}
\newcommand{\bk}{\mathbf{k}}

\title[Bias renormalisation]{Renormalisation of linear halo bias in N-body simulations}

\author[Werner \& Porciani]{
Kim F. Werner,$^{1}$
Cristiano Porciani$^{1}$\thanks{E-mail:
kwerner@astro.uni-bonn.de,\newline porciani@astro.uni-bonn.de}
\\
$^{1}$Argelander-Institut f\"ur Astronomie, Auf dem H\"ugel 71, 53121 Bonn\\
}

\date{Accepted XXX. Received YYY; in original form ZZZ}

\pubyear{2019}

\begin{document}
\label{firstpage}
\pagerange{\pageref{firstpage}--\pageref{lastpage}}
\maketitle

\begin{abstract}
The interpretation of redshift surveys requires modeling the relationship between 
large-scale fluctuations in the observed number density of tracers, $\delta_\mathrm{h}$, and the underlying matter density, $\delta$.
Bias models often express $\delta_\mathrm{h}$ as a truncated series of integro-differential operators acting on $\delta$, each weighted by a bias parameter. Due to the presence of `composite operators' (obtained by multiplying fields evaluated at the same spatial location), the linear bias parameter measured from clustering statistics does not coincide with that appearing in the bias expansion. 
This issue can be cured by re-writing the expansion in terms of `renormalised' operators. After providing a pedagogical and comprehensive review of bias renormalisation in perturbation theory,
we generalize the concept to non-perturbative dynamics and successfully apply it to dark-matter haloes extracted from a large suite of N-body simulations. 
When comparing numerical and perturbative results, we highlight the effect of the window function employed to smooth the random fields.
We then measure the bias parameters as a function of halo mass by fitting a non-perturbative bias model
(both before and after applying renormalisation)
to the cross spectrum $P_{\delta_\mathrm{h}\delta}(k)$. Finally, we employ Bayesian model selection to determine the optimal operator set to describe $P_{\delta_\mathrm{h}\delta}(k)$ for $k<0.2\,h$ Mpc$^{-1}$ at redshift $z=0$.
We find that it includes $\delta, \nabla^2\delta, \delta^2$ and the square of the traceless tidal tensor, $s^2$. Considering higher-order terms (in $\delta$) leads to overfitting as they cannot be precisely constrained by our data. We also notice that next-to-leading-order perturbative solutions are inaccurate for $k\gtrsim 0.1$ \hmpc.
\end{abstract}




\section{Introduction}
\label{sec:intro}

The idea that galaxies could be biased tracers of the underlying matter distribution dates back to the 1980s and is intimately linked with the development of the cold-dark-matter (CDM\footnote{Table~\ref{tab:abbrv} lists all the acronyms used in the paper.}) cosmological model \citep[e.g.][]{kaiser1984spatial,davis1985evolution,rees1985mechanisms,dekel1987physical,white1987galaxy}. For many years, the leading galaxy-biasing model assumed a local and deterministic relation between the smoothed galaxy density contrast $\delta_\mathrm{g}$  and the matter density contrast $\delta$ evaluated at the same position ${\mathbf x}$, i.e. $\delta_{\mathrm g}(\mathbf x)=f[\delta(\mathbf x)]$ where $f$ denotes a generic `bias function' \citep{fry1993biasing}. The freedom in the function $f$ was often described in terms of the `bias parameters' i.e. the Taylor coefficients $b_i$ appearing in the `bias expansion' $\delta_{\mathrm g}=b_0+b_1 \,\delta+b_2\,\delta^2+\dots$.
These parameters depend on the coarse-graining scale used to define $\delta$ and $\delta_{\mathrm g}$. The reasons for smoothing are multiple: (i) galaxies are discrete objects and we want to define a continuous density field; (ii) our model aims at describing the largest scales only; (iii) the bias expansion should be well-behaved so that it can be truncated at finite order; (iv) the dynamical model we use to evolve cosmological perturbations breaks down on small scales. 

This bias model is applicable to any tracers of the large-scale structure (LSS) of the Universe (e.g. galaxy clusters, dark-matter (DM) haloes, galaxies detected with different selection criteria) by picking the appropriate function $f$ or, equivalently, the corresponding set of bias parameters. In particular, if one considers DM haloes, the local ansatz for the bias relation is supported by spherically-symmetric models of gravitational collapse which also provide predictions for the bias parameters as a function of halo mass \citep[e.g.][]{mo1996analytic, mo1997high, porciani1998excursion}.

It was later realised that anisotropic gravitational collapse generates non-local and stochastic terms in the bias relation \citep{catelan1998bias} and that this originates leading-order (LO) corrections to the galaxy bispectrum \citep{catelan2000two}. The implications of this result were fully appreciated only a decade later when local bias was shown to be insufficient to describe the clustering of DM haloes \citep{manera2011local, matsubara2011nonlinear, roth2011testing, pollack2012modelling, pollack2014new} and evidence for bias corrections that are quadratic in the tidal tensor was found in N-body simulations
\citep{baldauf2012evidence,chan2012gravity}. At the same time, it was realized that the bias expansion should also contain a series of derivative
terms starting with $R^2\,\nabla^2 \delta$ because haloes and galaxies collect material from an extended region of space of characteristic size $R$ \citep[e.g.][]{desjacques2008baryon,mcdonald2009clustering, desjacques2010modeling,schmidt2013peak,fujita2016very}. The current consensus is that the bias relation should be compatible with all possible symmetries of gravitational instability \citep{mcdonald2009clustering, kehagias2014consequences,senatore2015bias,eggemeier2018bias} and thus depend on several other fields than $\delta$ (see section~\ref{sec:biaseft}).

Different techniques have been developed to measure the corresponding bias coefficients for DM haloes in numerical simulations. One possibility is to fit model predictions to the N-body data for various clustering statistics -- e.g. power spectra or two-point correlation functions -- \citep{saito2014understanding, bel2015non, hoffmann2015comparing, hoffmann2017linear, modi2017halo, hoffmann2018testing} or even at the field
level \citep{roth2011testing, schmittfull2018modeling}.
Alternatively, one can measure the response of the halo population to long-wavelength perturbations in the matter density \citep{li2016separate, baldauf2016linear, lazeyras2016precision, lazeyras2019robust}.
Lastly, bias parameters can be obtained
by cross-correlating the corresponding fields that appear in the bias expansion with the halo density distribution \citep{abidi2018cubic,lazeyras2018beyond}.

Another important line of research has examined how measurable clustering statistics of tracers depend on the bias relation and on the statistical properties of the underlying matter-density field. This is, in fact, key to interpret results from galaxy redshift surveys. For a local bias relation and in the presence of a Gaussian matter-density field, the ratio $b^2_{\mathrm{eff}}(k)$ between the (shot-noise subtracted) tracer- and matter power spectra tends to the constant $b_1^2$ when $k\to 0$ \citep{szalay1988constraints,coles1993galaxy,fry1993biasing,gaztanaga1998testing}. However, for a non-Gaussian matter density distribution, the constant does not coincide with the linear bias parameter and also depends on the cumulants of the matter density as well as higher-order bias parameters \citep{scherrer1998constraints}. In particular, if the statistics of the matter distribution are derived using standard perturbation theory, the LO correction to $b_1^2$ scales with the variance of the linear density field that grows large when the coarse-graining scale becomes small
\citep[][see also our section~\ref{sec:COSPT} for further details]{heavens1998non}. This sensitivity of the theoretical predictions on the smoothing scale that defines the bias function complicates the use of bias models to interpret observational data.

Inspired by applications of the renormalisation group to the perturbative solution of ordinary differential equations, \citet{mcdonald2006clustering} proposed a different interpretation of the theoretical predictions. The key concept is that the parameters $b_i$ appearing in the bias expansion do not actually correspond to the physical constants that can be measured in a survey. They are, instead, bare quantities that do not take into account the contribution of perturbative (loop) corrections to the leading terms. One can then rewrite the result of the perturbative calculations for the power spectrum of the tracers in terms of measurable `renormalised\footnote{The whole procedure parallels Wilsonian renormalisation in field theories. The coarse-graining scale of the density fields here replaces the cutoff that regularises the loop integrals in the quantum field theories.}' quantities that do not depend on the smoothing scale. \citet{mcdonald2009clustering} extended this approach to a bias relation that also depends on the tidal field. More recently, by using the diagrammatic representation of perturbation theory (PT), \citet{assassi2014renormalized} derived a set of conditions that the bias expansion must satisfy to be perturbatively renormalised order by order \citep[see also][]{senatore2015bias}. From these studies it emerges that only bias expansions that include terms proportional to the tidal field and to derivative of the matter density (e.g. $\nabla^2 \delta$) can be renormalised.

Bias renormalisation is analogous to methods used in statistical field theory and quantum field theory. In general, these techniques provide a convenient way to build `effective theories' that describe physics at a given range of length (and/or time, mass, etc.) scales and with a given accuracy by using a finite set of parameters. This bypasses the need for the `full theory' that includes physics on all scales but might be intractable or, even, unknown. Likewise, the goal of bias renormalisation is to build an effective model for the clustering statistics of tracers. Much of the jargon used in the literature on bias renormalisation is imported from other branches of physics. Less theoretically inclined researchers shy away from the unfamiliar notation and concepts. One of the motivations of this paper is to provide a pedagogical review of the subject and some concrete examples to see it `in action'. A second one is that there are a number of questions that are still unanswered. Are all the possible operators allowed by symmetries necessary in the bias expansion? Down to what distances does the renormalised bias expansion (truncated to some order) give an accurate description of the observed clustering statistics of tracers like galaxies?

Although the renormalisation of the bias expansion has always been discussed within the context of cosmological PT, only some of its aspects are fully perturbative. If one writes down a bias expansion in terms of Taylor coefficients, renormalisation is necessary also when one deals with the exact dynamics (extracted e.g. from simulations as in our case) in order to account for the non-linear terms (e.g. $\delta^2$) that generate smoothing-dependent spectra. In general, renormalisation fixes a `language' problem in the way we describe biasing, and PT provides an approximate solution on how to implement the remedy in practice. Here we present a first step towards implementing renormalisation within a large suite of N-body simulations where the dynamics is non-perturbative. This will shed new light on the limitations of the perturbative approach.

Usually, tests of the renormalised bias models are conducted by fitting the full perturbative expression for some clustering statistics to N-body simulations \citep[e.g.][]{saito2014understanding}.
We instead apply the renormalisation procedure step by step to the numerical data, examining all the contributions due to the relevant fields separately. For simplicity, we focus on the cross power spectral density $P_{\delta_\mathrm{h}\delta}(k)$ of the matter and halo density fields\footnote{With a little abuse of notation, we denote by $f(\mathbf{k})$ the Fourier transform of the function $f(\mathbf{x})$ where $\mathbf{x}$ indicates the comoving position.} defined as $\langle \delta_{\mathrm h}({\mathbf k})\,\delta({\mathbf k}') \rangle=(2\pi)^3\,P_{\delta_\mathrm{h}\delta}(k)\,\delta_{\mathrm D}({\mathbf k}+{\mathbf k}')$, where the brackets $\langle \dots \rangle$ denote the average over an ensemble of realizations and $\delta_\mathrm{D}(\mathbf{k})$ is the Dirac delta distribution in three dimensions.

The detailed goals of this work are to
(i) apply the  Wilsonian renormalisation-group (RG) method to halo bias in a large suite of N-body simulations and study the behaviour of the fields appearing in the bias expansion as a function of the coarse-graining scale; 
(ii) measure the bias parameters of the DM haloes extracted from simulations by fitting the halo-matter cross-power spectrum $P_{\delta_\mathrm{h}\delta}(k)$;
(iii) follow their evolution (running) induced by the RG coarse-graining scale;
(iv) test that they can be renormalised by re-arranging the terms in the bias expansion;
(v) use Bayesian model-selection techniques to determine which bias parameters are necessary to accurately describe $P_{\delta_\mathrm{h}\delta}(k)$ up to $k=0.2$ \hmpc.
In this paper, we introduce the main concepts of our study and focus on the renormalisation of the linear bias coefficient, $b_1$. We plan to
discuss the renormalisation of the bias parameters at second order in our future work.

The structure of this paper is as follows: The theoretical motivations and methods of bias renormalisation are reviewed in section~\ref{sec:theobias} where we also present an original discussion on the impact of filter functions. The numerical techniques and N-body simulations we use are described in section~\ref{sec:numerics}. Our results on renormalisation are presented in section~\ref{sec:simrenorm} and the measurements of the bias parameter in section~\ref{sec:measureb}. Finally, our conclusions are laid down in section~\ref{sec:concl}.

\begin{table}
	\centering
	\caption{Acronyms used in the paper and sections or equations where they are introduced.}
	\label{tab:abbrv}
	\begin{tabular}{ccc}
		\hline
		Acronym & Meaning & Section \\
		\hline
		CDM & Cold dark matter & 1 \\
		DM & Dark matter & 1 \\
		LSS & Large-scale structure & 1 \\
		LO & Leading order & 1 \\
		PT & Perturbation theory & 1 \\
		RG & Renormalisation group & 1 \\
		SPT & Standard perturbation theory & 2.1 \\
		NLO & Next-to-leading order & 2.1 \\
		EFT & Effective field theory & 2.2 \\
		CIC & Cloud in cell & 3.2 \\
		FFT & Fast Fourier transform & 3.2 \\
		MCMC & Markov Chain Monte Carlo & 5.1\\
		WAIC & Widely applicable information criterion & 5.2,
		Eq.~(\ref{eq:waic}) \\
		AIC & Akaike information criterion & 5.2 \\ 
		DIC & Deviance information criterion & 5.2 \\
		NR & No renormalisation& 5.3, Eq.~(\ref{eq:biasexpansion}) \\
		IE & Intermediate expansion & 5.4, Eq.~(\ref{eq:biasexpansionrenshift}) \\
		RL & Linear renormalisation & 5.4, Eq.~(\ref{eq:biasexpansionRL}) \\
		RNL & Non-linear renormalisation & 5.4, Eq.~(\ref{eq:biasexpansionRNL}) \\
		\hline
	\end{tabular}
\end{table}

\section{Tracer bias in cosmological perturbation theory}
\label{sec:theobias}

\subsection{Standard perturbation theory in a nutshell}
Standard perturbation theory \citep[SPT, for a review, see][]{bernardeau2002large} describes matter as a pressureless and inviscid fluid to model the growth of density and velocity perturbations in a Friedmann-Robertson-Walker background with expansion factor $a$. The system formed by the continuity, Euler and Poisson equations is solved perturbatively. At any given time, the fastest growing solution for the density contrast, $\delta(\mathbf x)$, is expanded as
\begin{equation}
\delta(\mathbf{x})=\delta_1(\mathbf{x})+\delta_2(\mathbf{x})+\delta_3(\mathbf{x})+\dots
\label{deltaexp}
\end{equation}
where $\delta_1$ denotes the growing-mode solution to the linearized set of equations and  $\delta_n={\mathcal O}(\delta_1^n)$. The time evolution of $\delta_1$ is governed by the linear growth factor $D(a)$, such that $\delta_1 \propto D$. Similarly, the fastest growing mode for the divergence of the peculiar velocity
$\nabla\cdot \mathbf{v}=-aHf\,\theta$
-- where $H=\dot{a}/a$ is the Hubble parameter (the dot indicates differentiation with respect to cosmic time) and $f={\mathrm d} \ln D/{\mathrm d} \ln a$ --
is written as
\begin{equation}
 \theta(\mathbf{x})= \theta_1(\mathbf{x})+\theta_2(\mathbf{x})+\theta_3(\mathbf{x})+\dots\;,
\label{thetaexp}
\end{equation}
with $\theta_1=\delta_1$.
In Fourier space, 
\begin{multline}
\delta_n \big(\mathbf{k} \big) = \int F_n \big(\mathbf{k}_1, \dots, \mathbf{k}_n \big) \, \delta_1 \big(\mathbf{k}_1 \big) \dots \delta_1 \big(\mathbf{k}_n \big)\\\delta_\mathrm{D} \big (\mathbf{k}_1 + \dots+\mathbf{k}_n - \mathbf{k} \big) \, \frac{\mathrm{d}^3k_1}{(2\pi)^3} \dots \frac{\mathrm{d}^3k_n}{(2\pi)^3}
\label{eq:delta2}
\end{multline}
and
\begin{multline}
\theta_n \big(\mathbf{k} \big) = \int G_n \big(\mathbf{k}_1, \dots, \mathbf{k}_n \big) \, \delta_1 \big(\mathbf{k}_1 \big) \dots \delta_1 \big(\mathbf{k}_n \big)\\\delta_\mathrm{D} \big (\mathbf{k}_1 + \dots+\mathbf{k}_n - \mathbf{k} \big) \, \frac{\mathrm{d}^3k_1}{(2\pi)^3} \dots \frac{\mathrm{d}^3k_n}{(2\pi)^3} \;,
\label{eq:theta2}
\end{multline}
where the kernels $F_n$ and $G_n$ are homogeneous
functions of degree zero that describe the couplings between Fourier modes generated by the dynamical non-linearities. These functions obey recursion relations that can be solved order by order starting from $F_1=G_1=1$ \citep[e.g.][]{goroff1986coupling}. For instance, the second-order kernels (symmetrized over permutations of their arguments) are
\begin{equation}
F_2\big( \mathbf{k}_1, \mathbf{k}_2)= \frac{5}{7} + \frac{1}{2}\, \frac{\mathbf{k}_1 \cdot \mathbf{k}_2}{k_1 k_2} \,\Bigg(\frac{k_1}{k_2} + \frac{k_2}{k_1}\Bigg) + \frac{2}{7}\, \Bigg(\frac{\mathbf{k}_1 \cdot \mathbf{k}_2}{k_1 k_2}\Bigg)^2 \;,
\end{equation}
\begin{equation}
G_2 \big(\mathbf{k}_1, \mathbf{k}_2)= \frac{3}{7} + \frac{1}{2}\, \frac{\mathbf{k}_1 \cdot \mathbf{k}_2}{k_1 k_2} \,\Bigg(\frac{k_1}{k_2} + \frac{k_2}{k_1}\Bigg) + \frac{4}{7}\, \Bigg(\frac{\mathbf{k}_1 \cdot \mathbf{k}_2}{k_1 k_2}\Bigg)^2\;,
\end{equation}
Although these expressions are exact only in an Einstein-de Sitter universe, they provide an excellent approximation also in the $\Lambda$CDM scenario \citep{bernardeau1994effects,lee2014exact}.

Assuming Gaussian initial conditions allows us to compute perturbative expansions for statistical quantities averaged over an ensemble of realisations of $\delta_1$. For instance, let us consider the power spectrum of matter density perturbations, $P_{\delta\delta}(k)$, defined as $\langle \delta(\mathbf{k})\, \delta(\mathbf{k}') \rangle=(2\pi)^3\,P_{\delta\delta}(k)\,\delta_{\mathrm D}(\mathbf{k}+\mathbf{k}')$. Within SPT and up to fourth order in $\delta_1$, we can write the `correlator' $\langle \delta \delta \rangle\simeq \langle\delta_1\delta_1 \rangle+\langle\delta_2\delta_2 \rangle + \langle\delta_3\delta_1 \rangle$ (the obvious dependence on the wavevectors is understood here to simplify notation). Therefore, $P_{\delta\delta}(k)$ can be approximated as 
\begin{equation}
P_{\delta\delta}(k)\simeq P^{(11)}_{\delta\delta}(k)+P^{(22)}_{\delta\delta}(k)+P^{(31)}_{\delta\delta}(k)\;,
\end{equation} 
where the LO term $P^{(11)}_{\delta\delta}(k)$ coincides with the linear power spectrum $P_{11}(k)$ while the next-to-leading-order (NLO) corrections are
\begin{align}
P^{(22)}_{\delta\delta}(k)&=2\,\int F_2^2(\mathbf{q}, \mathbf{k}-\mathbf{q})\, P_{11}(|\mathbf{k}-\mathbf{q}) |)\,P_{11}(q)\,\frac{\mathrm{d}^3q}{(2\pi)^3}\;,\\
P^{(31)}_{\delta\delta}(k)&=3\, P_{11}(k) \int F_3(\mathbf{q}, \mathbf{-q}, \mathbf{k}) \,P_{11}(q)\, \frac{\mathrm{d}^3q}{(2\pi)^3}\;.
\end{align}
A powerful diagrammatic technique has been introduced to conveniently perform the SPT expansion of ensemble-averaged statistics \citep[e.g.][]{bernardeau2002large}. This is analogous to the method introduced by Feynman in quantum electrodynamics. In most cases, LO contributions are associated with tree diagrams (in the sense of graph theory) and are thus called `tree-level terms'. Evaluating these quantities does not require any integration (see e.g. $P_{11}(k)$ above). On the other hand, terms associated with diagrams containing $n$-loops give rise to `$n$-loop corrections' that require $n$ integrations (see e.g. the 1-loop term $P^{(22)}_{\delta\delta}(k)+P^{(13)}_{\delta\delta}(k)$). Note that, if the tree-level term vanishes, the LO is given by the 1-loop terms (see e.g. section~\ref{sec:COSPT}).

SPT mainly suffers from two limitations. First, being a perturbative technique, it is expected to break down when and where $|\delta(\mathbf{x})| \simeq 1$. Second, the gravitational collapse of an initially cold distribution of collisionless DM develops multi-stream regions where the velocity field is not single-valued. However, the pressureless-fluid approximation adopted by SPT does not account for this phenomenon which alters the dynamics of the system. N-body simulations show that, within the $\Lambda$CDM scenario, SPT provides rather accurate predictions for 2- and 3-point statistics at redshifts $z>1$ and $k\lesssim 0.2\ h$ Mpc$^{-1}$ while it becomes increasingly imprecise at lower redshifts (for the same wavenumbers) as the variance of the density perturbations approaches unity \citep[e.g.][]{carlson2009critical, nishimichi2009modeling, blas2014cosmological}.

\subsection{Biasing as an effective field theory}
\label{sec:biaseft}
We are interested in describing how discrete objects (for example DM haloes or galaxies) trace the smooth matter density field throughout the Universe. Since haloes assemble from the gravitational collapse of matter on small scales and galaxies form within them, it is reasonable to assume that some deterministic relationship exists between the density of matter and that of the discrete tracers on large scales. Numerical simulations and simple toy models that associate haloes to peaks in the initial density field provide supporting evidence in favour of this argument. However, the physics of halo and galaxy formation is complex, highly non-linear and non-perturbative. Although we can simulate it with the help of a computer, we are not able to make analytical predictions. Given these circumstances, for practical applications like the interpretation of galaxy redshift surveys, it makes sense to opt for a simplified description that holds true only on large scales.

Analogous problems in other fields of physics led to the
development of effective field theories (EFTs). We highlight here how the basic concepts of an EFT can be used to model large-scale biasing. We first introduce the effective overdensity field of the tracers\footnote{The subscript `h' refers to DM haloes, but the framework is applicable to any tracer.}, $\deh({\mathbf x},t)$, which describes their large-scale clustering, but is blind to their precise distribution on small scales. In order to relate this quantity to the underlying 
mass-overdensity field, $\delta$, it is necessary to identify all possible dependencies that are compatible with the symmetries of the problem. Assuming that the tracers are non-relativistic implies that gravitational physics is fully described in terms of the (rescaled) peculiar gravitational potential $\phi$, defined so that $\nabla^2 \phi=\delta$. Taking into account the equivalence principle and that $\deh$ is a scalar under rotations, it follows that $\deh$ can only depend on scalar combinations of second spatial derivatives of the peculiar gravitational potential
$\partial_i\partial_j \phi$, and first spatial derivatives of the peculiar velocity field $\partial_j {\varv}^i$
\citep{mcdonald2009clustering, kehagias2014consequences, senatore2015bias}.

The second key step in the construction of an EFT for biasing is to consider that the physics regulating the clustering of tracers is non-local in space and time. The material that forms a tracer was dispersed within an extended patch of characteristic size $R$ at early times and needed a characteristic time $T$ to assemble together. In general, $R\lesssim 10$ \mpch\ (the Lagrangian size of galaxy clusters) which is small compared with the scales we want to describe using our EFT. On the other hand, $T$ is never short compared to the Hubble time. It follows that the effective theory of biasing should be approximately local in space but non-local in time \citep{senatore2015bias}. In mathematical terms, this means that $\deh$ should depend on the cosmological perturbations evaluated along the past worldlines of the fluid elements that end up forming the tracers at a given location (or, better, within a given patch).

Since we are only interested in the large spatial scales, the third step is to expand the generic (and unknown) functional $\deh$ in powers of the cosmological fluctuations and their spatial derivatives (to account for the mild spatial non-locality). Note that the tensor $\partial_i\partial_j \phi$ is a dimensionless quantity and its spatial derivatives need to be multiplied by a length scale of order $R$ in the expansion. It follows that all derivative corrections will be suppressed on scales much larger than $R$. This is more easily seen in Fourier space. For instance, let us consider the lowest derivative correction to terms that are proportional to $\delta$, i.e. $\sim R^2 \nabla^2 \delta$ (as we need a scalar under rotations). In Fourier space this term is proportional to $(kR)^2$ and thus heavily suppressed for scales $k\ll R^{-1}$. On the other hand, this also implies that the derivative expansion breaks down on scales of order $R$.

Finally, the non-locality in time is accounted for by making some further hypotheses, namely by assuming that cosmological perturbations evolve as in SPT. In this case, the expansion of $\deh$ is re-organised to avoid duplication of terms (for instance the velocity divergence coincides with the overdensity at linear order) and its coefficients are re-defined as integrals over time of the original ones \citep{mirbabayi2015biased, angulo2015one, desjacques2016large, desjacques2018galaxy}.

Up to third order in the cosmological perturbations, one thus obtains
\begin{align}
\delta_\mathrm{h}= b_0+b_1 \delta&+
b_{\nabla^2\delta} \nabla^2 \delta+
b_2 \delta^2 + b_{s^2} s^2 \nonumber\\
&+b_3 \delta^3+b_{\delta s^2} \delta s^2+ b_{s^3} s^3+ b_{\Gamma_3} \Gamma_3
\;,
\label{eq:biastidal}
\end{align}
where we have
(i) decomposed the tensor $\partial_i \partial_j \phi$ into its trace $\delta$ and the traceless part,
\begin{equation}
s_{ij}=\partial_i \partial_j \phi- \frac{1}{3} \delta_{ij}^{\mathrm{K}}\, \delta= \left(\partial_i \partial_j\nabla^{-2}- \frac{1}{3}\delta_{ij}^{\mathrm{K}}\right)\delta = \gamma_{ij}\, \delta\;,
\label{eq:sij}
\end{equation}
(with $\nabla^{-2}$ the inverse of the Laplacian operator and $\delta_{ij}^{\mathrm{K}}$ the Kronecker symbol);
(ii) introduced the scalars $s^2=s_{ij}s_{ji}$, 
$s^3=s_{ij}s_{jk}s_{ki}$;
(iii) used the (rescaled) velocity potential $\phi_\varv$ such that $\nabla^2 \phi_\varv=\theta$ to define the tensor $p_{ij}=\partial_i \partial_j \phi_\varv-\frac{1}{3} \delta_{ij}^{\mathrm{K}} \theta$ and the operator\footnote{The bias expansion is sometimes written using the so-called second- and third-order `Galileon' operators, $\mathcal{G}_2(\phi)$ and $\mathcal{G}_3(\phi)$
\citep[e.g.][see also \citealt{chan2012gravity}]{assassi2014renormalized}. The relation between this set of operators and ours is
$\mathcal{G}_2(\phi)=s^2-(2/3)\,\delta^2$ and $\mathcal{G}_3(\phi)=-\delta^3/9- s^3+\delta s^2/2$.
Note that $\Gamma_3=\mathcal{G}_2(\phi)-\mathcal{G}_2(\phi_\varv)$.}
$\Gamma_3=s^2-p^2-(2/3)(\delta^2-\theta^2)$
with $p^2=p_{ij}p_{ji}$;
(iv) included only the leading higher-derivative term $\propto \nabla^2 \delta$, for simplicity.

In compact form, the bias expansion can be written as
\begin{equation}
\delta_{\mathrm h}=\sum_{O} b_{O}\,{O}\;,
\label{eq:exp}
\end{equation}
where the sum runs over a fixed basis of operators, ${O}$, that are compatible with the symmetries and the evolution of cosmological perturbations while the bias parameters, $b_{O}$, depend on the characteristics of the population of tracers (e.g. the halo mass, the galaxy luminosity or the intensity of a particular emission line).

The expressions above are meant to relate the spatial distribution of tracers on large scales with the underlying long-wavelength cosmological perturbations in a deterministic way. However, the short-wavelength fluctuations (to which our effective theory is blind by construction) also play a role in determining the precise location of the tracers. In order to account for this phenomenon in the theory, we introduce a certain degree of randomness in the bias expansion by assuming that the bias coefficients have also a zero-mean stochastic component $\epsilon_{O}$, i.e.
\begin{equation}
\delta_{\mathrm h}=\sum_{O} \left(b_{O}+\epsilon_{O}\right)\,{O}\;.
\label{eq:biasstoc}
\end{equation}
Furthermore, it is customary to treat each stochastic term as a perturbation ${\mathcal O}(\delta)$. For example, to second order, we can write:
\begin{equation}
\delta_\mathrm{h}= b_0+\epsilon_0+b_1 \delta + b_{\nabla^2 \delta} \nabla^2 \delta+\epsilon_1\delta+ \epsilon_{\nabla^2 \delta} \nabla^2 \delta +b_2 \delta^2 + b_{s^2} s^2\;.
\end{equation}
Under the assumption that the stochastic terms do not correlate
with the long-wavelength cosmological perturbations, their statistical properties are fully determined by their auto- and cross-correlation functions (or auto- and cross-spectra in Fourier space). In the literature, it is often assumed that the cross-spectrum between $\epsilon_{O}$ and $\epsilon_{O'}$ can be written as a series expansion in $k^2$,
\begin{equation}
    \Upsilon_{O O'}(k)=\Upsilon_{0, O O'}+\Upsilon_{2, O O'} k^2+ \Upsilon_{4, O O'} k^4+\dots \;.
\end{equation}
Once again, this reflects the fact that $\delta_{\mathrm h}$ should be determined by the value assumed by the stochastic fields within an extended region of space and not only at one point. However, when taken term by term, this expansion makes little sense in configuration space. While $\Upsilon_{0, O O'}$ gives rise to a standard shot-noise term with a 2-point correlation that is proportional to the Dirac delta distribution at zero lag, $\delta_{\mathrm D}(r)$, the remaining terms generate contributions proportional to $\nabla^{2n} \delta_{\mathrm D}(r)$. These `contact terms' (as they are generally dubbed in field theory) appear because we are considering an asymptotic expansion in Fourier space that breaks down on small scales. A more sophisticated treatment of the short-wavelength modes should then be used to discuss the statistics of tracers in configuration space.

A note is in order regarding equations (\ref{eq:biastidal}), (\ref{eq:exp}), and (\ref{eq:biasstoc}). Implicit in the definition of all fields is a low-pass smoothing procedure that isolates the long-wavelength modes to which the theory applies. In particular, it is necessary to specify how the smoothing is done for the non-linear terms like $\delta^2$ and $s^2$. In fact, the long-wavelength modes of $\delta^2$ depend on the short-wavelength modes of $\delta$ implying that $\delta^2$ cannot be reliably computed with a perturbative approach and that it can grow large even at low $k$. Since we want to consider only the contribution of the large-scale modes of the density perturbations, the only meaningful interpretation of the r.h.s. in equations (\ref{eq:biastidal}), (\ref{eq:exp}), and (\ref{eq:biasstoc}) is to smooth the matter overdensity field first and then use the low-pass filtered $\delta$ to evaluate the non-linear operators \citep{heavens1998non}. Filtering out the short-wavelength modes with $k>\Lambda$ is thus necessary for several reasons: (i) as we have already mentioned, to only consider the scales that are described by the theory; (ii) to ensure that $|\delta (\mathbf{x})|\ll 1$ and a series expansion of the bias relation in the cosmological perturbations makes sense; and (iii) to be able to apply SPT. 

The ultimate goal of the effective theory of biasing is to model the observed $n$-point correlation functions (and their corresponding multi-spectra in Fourier space) for tracers on large scales. In this work, we focus on two-point statistics in Fourier space. The model predictions are obtained by correlating the Fourier transform of equation (\ref{eq:biasstoc}) either with itself or with the Fourier transform of the overdensity field. This provides a systematic expansion of the spectra organized according to the importance of its terms based on a power-counting argument. The underlying assumption is that there exist two well-defined scales: a non-linearity scale $R_{\mathrm{nl}}$ (at which physics becomes non-perturbative and the expansion in powers of the cosmological fluctuations breaks down) and a non-locality scale $L$ (at which the expansion in powers of the higher-derivative terms breaks down). The different contributions to the spectra scale as powers of $kL$ and $kR_{\mathrm{nl}}$. Truncating the expansion at a given order thus provides results with reliable error estimates without referencing any quantity outside of the theory. The bias parameters and the coefficients that define the statistical properties of the stochastic fields are free parameters that need to be tuned in order to match observations. However, since the non-linear operators of the theory are heavily dependent on the smoothing scale, the best-fitting coefficients will inherit a dependence on $\Lambda$. As we have already mentioned in the introduction, an analogy can be established between this description of galaxy biasing and the Wilsonian renormalisation in quantum field theory. In brief, the bias series can be re-organized so that to eliminate the so-called `UV-sensitive' terms that depend on $\Lambda$ order by order in the perturbative expansion. This is equivalent to expressing the bias relation in terms of a set of `renormalised' operators $[{O}]$ built from the original operator basis $O$
\begin{equation}
\delta_{\mathrm h}=\sum_{[O]} (b_{[O]}+\epsilon_{[O]})\,[O]\;.
\end{equation}

\subsection{Bias renormalisation}
\label{sec:renormintro}
The bias renormalisation forms the main subject of our investigation and, for this reason, we explore the related concepts in more detail.

\subsubsection{UV-sensitivity of composite operators}
The bias expansion given in equation (\ref{eq:biasstoc}) contains
operators obtained by multiplying two fields evaluated at the same spatial location, like $\delta^2$ and $s^2$. These `local operators' or `composite operators' (as they are generally called in field theories) are very sensitive to the short-wavelength modes of $\delta$ that are not accurately modelled by SPT. This can be easily understood, for example, by inspecting the Fourier transform of $\delta^2$:
\begin{equation}
    \delta^2({\mathbf k})=\int \delta({\mathbf q})\,\delta({\mathbf k}-{\mathbf q})\, \frac{\mathrm{d}^3q}{(2\pi)^3}\;.
\end{equation}
Even for small values of $k$, $\delta^2$ receives contributions
from all scales. For instance, its expectation value,
\begin{equation}
\langle \delta^2({\mathbf x})\rangle= \frac{1}{2\pi^2}\int_0^\infty q^2\, P_{\delta\delta}(q)\, {\mathrm d}q\equiv \sigma^2\;,
\label{variance}
\end{equation}
can get very large (or even diverge) if the density perturbations have substantial power at small scales. Similarly, the cross spectrum of $\delta^2$ with the matter-density fluctuations, defined as $\langle \delta^2({\mathbf k})\, \delta({\mathbf k}')\rangle=(2\pi)^3\,P_{\delta^2 \delta}(k)\,\delta_{\mathrm D}(\mathbf{k}+\mathbf{k}')$, coincides with an integral over all scales of the matter bispectrum 
\begin{equation}
P_{\delta^2 \delta}(k)= \int B_{\delta\delta\delta}({\mathbf q},{\mathbf k}-{\mathbf q},-{\mathbf k}) \, \frac{\mathrm{d}^3q}{(2\pi)^3}\;,
\label{eq:crossd2d}
\end{equation}
where $\langle \delta({\mathbf p})\, \delta({\mathbf q})\, \delta({\mathbf k}) \rangle=(2\pi)^3 \,B_{\delta\delta\delta}(p,q,k) \,\delta_{\mathrm D}({\mathbf p}+{\mathbf q}+ {\mathbf k})$. All this implies that, if a low-pass filter $W(k,\Lambda)$ is applied to $\delta$, then $\delta^2(\bk)$ inherits a strong dependence on $\Lambda$. In brief, we say that the composite operators $\delta^2$ and $s^2$ are UV-sensitive.

\subsubsection{Composite operators in SPT}
\label{sec:COSPT}
To LO in SPT, we can write the matter power spectrum $P_{\delta\delta}^{\mathrm{(LO)}}(k)=P_{11}(k)$ and the bispectrum $B_{\delta\delta\delta}^{\mathrm{(LO)}}(k_1,k_2,k_3)=B^{(112)}_{\delta\delta\delta}(k_1,k_2,k_3)+B^{(121)}_{\delta\delta\delta}(k_1,k_2,k_3)+B^{(211)}_{\delta\delta\delta}(k_1,k_2,k_3)$. By substituting these approximations in equations (\ref{variance}) and (\ref{eq:crossd2d}) one obtains
\begin{equation}
\sigma^2_{\mathrm{(LO)}}= \lim_{\Lambda \to \infty} \frac{1}{2\pi^2}\int_0^\Lambda q^2\, 
P_{11}(q)\, {\mathrm d}q\equiv \lim_{\Lambda \to \infty} \sigma_1^2(\Lambda)\;,
\label{eq:variancelo}
\end{equation}
and
\begin{align}
P^{\mathrm{(LO)}}_{\delta^2 \delta}(k)&= \int \left[B^{(112)}_{\delta\delta\delta}({\mathbf q},{\mathbf k}-{\mathbf q},-{\mathbf k})+\mathrm{cyclical}\right] \, \frac{\mathrm{d}^3q}{(2\pi)^3}\nonumber \\&=
P_{\delta^2 \delta}^{(22)}(k) + P_{\delta^2 \delta}^{(31)}(k) \;,
\label{eq:crossd2dlo}
\end{align}
with 
\begin{equation}
P_{\delta^2 \delta}^{(22)}(k) =2 \int F_2\big(\mathbf{q}, \mathbf{k}-\mathbf{q}\big) P_{11}\big(q\big)\, P_{11}\big(|\mathbf{k} - \mathbf{q}|\big) \frac{\mathrm{d}^3q}{(2 \pi)^3} \, ,
\label{eq:P22}
\end{equation}
and
\begin{equation}
P_{\delta^2 \delta}^{(31)}(k) =4 P_{11}\big(k\big) \int F_2\big(-\mathbf{q}, \mathbf{k}\big) P_{11}\big(q\big) \frac{\mathrm{d}^3q}{(2 \pi)^3} \, .
\label{eq:P31}
\end{equation}
The behaviour of these cross-spectra at large scales ($k\to 0$) is usually determined by using spherical coordinates (with polar axis $\mathbf{k}$) and limiting the integration in $q$ with an upper cutoff $\Lambda$. It turns out that
\begin{equation}
P_{\delta^2 \delta}^{(31)}(k)=
\frac{68}{21}\sigma_1^2(\Lambda)\,P_{11}(k)\,,  
\label{eq:P6821}
\end{equation}
which shows that the normalisation of $P_{\delta^2 \delta}^{(31)}(k)$ depends on $\Lambda$ \citep{heavens1998non}. This is an example of a UV-sensitive term. On the other hand, the $k\to 0$ limit of $P_{\delta^2 \delta}^{(22)}(k)$ does not show any cutoff dependence.

\subsubsection{Renormalisation and counterterms}
Fur illustrative purposes only, let us now consider a simplified bias expansion such that 
\begin{equation}
\delta_\mathrm{h}(\mathbf{x})= b_0+b_1 \delta(\mathbf{x})+b_2 \delta^2(\mathbf{x})
\label{simplebias}
\end{equation}
and evaluate the expectation value of $\delta_{\mathrm h}$ at NLO in SPT.

It is straightforward to obtain that $\langle \delta_\mathrm{h}(\mathbf{x})\rangle=b_0+b_2 \sigma_1^2(\Lambda)$. Since $\langle \delta_{\mathrm h}\rangle$ is an observable which is identically equal to zero, if we want to identify our truncated perturbative result with the actual measurement, it is necessary to set $b_0+b_2 \sigma_1^2(\Lambda)=b_0^{\mathrm{R}}=0$. This leads to the modified bias expansion $\delta_\mathrm{h}(\mathbf{x})=b_0^{\mathrm{R}}+ 
b_1 \delta(\mathbf{x})+b_2 
[\delta^2(\mathbf{x})-\sigma_1^2(\Lambda)]=
b_1 \delta(\mathbf{x})+b_2 [\delta^2(\mathbf{x})-\sigma_1^2(\Lambda)]$.

Starting from this expression, we now derive the cross spectrum between the tracer density field and the matter density to 1-loop in SPT. The tree-level result is $b_1 P_{11}(k)$, while the 1-loop corrections give $b_2 \big[P_{\delta^2 \delta}^{(22)}(k)+P_{\delta^2 \delta}^{(31)}(k)\big]$. Equation (\ref{eq:P6821}) shows that the loop corrections contain a term which coincides with the tree-level result rescaled by a cutoff-dependent coefficient. Remarkably, the combination
$[b_1+b_2\, (68/21)\, \sigma_1^2(\Lambda)]\, P_{11}(k)$ gives the dominant contribution when $k\to 0$. Inspired by Wilsonian renormalisation of field theories where the UV divergences are cancelled by a redefinition of the parameters of the theory, 
\citet{mcdonald2006clustering} proposes to replace $b_1+b_2\, (68/21)\, \sigma_1^2(\Lambda)$ with the new (`renormalised' or `observable') linear bias coefficient $b_1^{\mathrm{R}}$. The resulting expression for the cross spectrum at one loop will thus be 
\begin{equation}
P_{\delta_\mathrm{h}\delta}(k)=b_1^\mathrm{R} P_{11}(k)+b_2\, P_{\delta^2 \delta}^{(22)}(k)\;.
\end{equation}
Here, $b_1^{\mathrm{R}}$ should be considered as a free parameter of the model that can be adjusted to fit observational data. On the other hand, the `bare' linear bias coefficient $b_1$ should be merely considered as a mathematical tool to perform calculations and should not be assigned any physical meaning. This technique can be iterated to renormalise higher-order bias parameters \citep{mcdonald2006clustering, mcdonald2009clustering}. The net effect of this procedure is that no UV-sensitive terms appear in the theoretical expressions for the ensemble averaged statistics of the tracers.

A different way to understand renormalisation is through the introduction of counterterms in the bias expansion. For instance, let us re-consider the example above. In order to eliminate the UV-sensitive term that appears in the NLO expression for $\langle \delta_{\mathrm h}(\mathrm{x})\rangle$, we could add the constant counterterm $\Delta b_0$ to the r.h.s. of equation (\ref{simplebias}) and get $\langle \delta_\mathrm{h}(\mathbf{x})\rangle=b_0+\Delta b_0+b_2 \sigma_1^2(\Lambda)$. By requiring that the expectation value of the theory coincides with the observed value (i.e. zero), we thus get $\Delta b_0=-b_0-b_2 \sigma_1^2(\Lambda)$. The new bias expansion (including the counterterm) can thus be re-organised as follows:
\begin{equation}
\delta_\mathrm{h}(\mathbf{x})=
b_1 \delta(\mathbf{x})+b_2 \left[\delta^2(\mathbf{x})\right]_0\;,
\end{equation}
where $\left[\delta^2(\mathbf{x})\right]_0\equiv \delta^2({\mathbf x})-\sigma_1^2(\Lambda)$ denotes a new operator compatible with the renormalisation of $b_0$. 

Next, we could add the counterterm $\Delta b_1 \delta(\mathbf{x})$ to the bias expansion in order to cancel out the UV-sensitive term in the NLO expression for the two-point statistic $P_{\delta_\mathrm{h}\delta}(k)$. In fact, the new term generates the correction $\Delta P_{\delta_\mathrm{h}\delta}(k)=\Delta b_1 P_{11}(k)$ and
by imposing that $b_1+\Delta b_1+ b_2 \,(68/21) \,\sigma_1^2(\Lambda)= b_1^{\mathrm{R}}$, we finally obtain 
\begin{equation}
\delta_\mathrm{h}(\mathbf{x})= b_1^{\mathrm{R}} \delta(\mathbf{x})+b_2 \left[\delta^2(\mathbf{x})\right]_1\;,
\end{equation}
where $\left[\delta^2(\mathbf{x})\right]_1\equiv \delta^2({\mathbf x})-\sigma_1^2(\Lambda)\,[1+(68/21)\, \delta(\mathbf{x})]$ represents the quadratic operator which is consistent with the renormalisation of both $b_0$ and $b_1$.

The examples above illustrate how the linear bias coefficient $b_1$ and the field $\delta^2$ can be consistently renormalised starting from the simplified model given in equation (\ref{simplebias}). Considering a more general bias expansion that includes additional composite operators requires further calculations. For instance, we might want to add a non-local bias term proportional to $s^2$. In this case, all the perturbative calculations we made to compute spectra and cross-spectra for $\delta^2$ can be easily generalised to $s^2$. By introducing the Fourier-space operator
\begin{equation}
S_2\left(\mathbf{k}_1, \mathbf{k}_2\right)= \gamma_{ij}(\mathbf{k}_1)\, \gamma_{ji}(\mathbf{k}_2)=\left( \frac{\mathbf{k}_1 \cdot \mathbf{k}_2}{k_1 k_2} \right)^2 -\frac{1}{3} \;,
\end{equation}
one finds that $\langle s^2({\mathbf x})\rangle=(4/3)\,\sigma^2$ and
\begin{equation}
P_{s^2 \delta}^{(31)}(k)=4 P_{11}\big(k\big) \int F_2\big(-\mathbf{q}, \mathbf{k}\big)\, S_2(\mathbf{q},\mathbf{k}-\mathbf{q})\, P_{11}\big(q\big) \frac{\mathrm{d}^3q}{(2 \pi)^3} \;,
\label{eq:P31s2}
\end{equation}
which is UV-sensitive as \citep{mcdonald2009clustering}
\begin{equation}
\lim_{k\to 0}\frac{
P_{s^2 \delta}^{(31)}(k)}{P_{11}(k)}=
\frac{136}{63}\sigma_1^2(\Lambda)\,.  
\label{eq:P13663}
\end{equation}
Thus, $s^2$ needs to be renormalised. Analogous considerations apply to other composite operators.

A systematic procedure for renormalising the bias expansion order by order in PT has been presented by \citet[][see also \citealt{senatore2015bias}]{assassi2014renormalized}. The renormalisation conditions for the generic operator ${O}$ appearing in the bias expansion are
\begin{equation}
    \langle [{O}]({\mathbf q})\, \delta_1({\mathbf q}_1) \dots \delta_1({\mathbf q}_m)\rangle=
    \langle {O}({\mathbf q})\, \delta_1({\mathbf q}_1) \dots \delta_1({\mathbf q}_m)\rangle^{\mathrm{(LO)}}\;,
\end{equation}
where ${\mathbf q}_i \to 0$ $\forall i$. In a diagrammatic representation, this means that the counterterms should be chosen so that to cancel the loop corrections obtained from diagrams in which different Fourier modes of $\delta$ that contribute to the operator ${O}$ are contracted among themselves as in equations (\ref{eq:P31}) and (\ref{eq:P31s2}).

\subsection{Cross-spectrum between matter and tracers}
\label{sec:model}
In this paper, we use a large suite of N-body simulations to test how accurately the third-order bias expansion given in equation (\ref{eq:biastidal}) describes the spatial distribution of biased tracers (namely, dark-matter haloes with different masses). In order to focus on the deterministic terms, we only consider the cross-spectrum between the matter-density field and the tracers, $P_{\delta_\mathrm{h}\delta}(k)$. 
As a reference, we provide here the perturbative result for this quantity (to NLO) expressed in terms of the renormalised linear bias parameter $b_1^\mathrm{R}$ \citep{mcdonald2009clustering,assassi2014renormalized,saito2014understanding,senatore2015bias,desjacques2016large}:
\begin{align}
P_{\delta_\mathrm{h}\delta}(k)&=(b_1^\mathrm{R}+  b_{\nabla^2\delta}k^2)\,[P_{\delta\delta}^{\mathrm{(LO)}}(k)+P_{\delta\delta}^{\mathrm{(NLO)}}(k)]
+ b_2\,P_{[\delta^2]_1 \delta}(k)\nonumber \\
&+ b_{s^2}\,P_{[s^2]_1 \delta}(k)
+ b_{\Gamma_3}P_{\Gamma_3 \delta}(k)
+\Upsilon_{2,\epsilon_1,\epsilon_{\mathrm m}}\,k^2\;.
\label{eq:fullpert}
\end{align}
In the expression above,
we have introduced the LO cross spectra of
the renormalised operators,
\begin{align}
P_{[\delta^2]_1 \delta}(k)= \lim_{\Lambda\to \infty} P_{\delta^2 \delta}^{(22)}(k)\;, 
\end{align}
\begin{align}
P_{[s^2]_1 \delta}(k)=\lim_{\Lambda\to \infty} \left[P_{s^2 \delta}^{(22)}(k)+B_{s^2\delta}(k)\right]\;, 
\end{align}
where\footnote{Equation (\ref{eq:P6821}) implies that $B_{\delta^2\delta}(k)=0$ as $P_{\delta^2 \delta}^{(31)}(k) \propto P_{11}(k)$.}
\begin{align}
P_{s^2 \delta}^{(22)}(k) = 2\int F_2(\mathbf{q}, \mathbf{k}-\mathbf{q})\,  S_2(\mathbf{q}, &\mathbf{k}-\mathbf{q})\, 
P_{11}(q) \nonumber \\ & \cdot P_{11}(|\mathbf{k}-\mathbf{q}|)\, \frac{\mathrm{d}^3q}{(2 \pi)^3}\;,
\end{align}
\begin{align}
B_{s^2\delta}(k)&=P_{s^2 \delta}^{(31)}(k)-\frac{136}{63}\sigma_1^2(\Lambda)\,P_{11}(k)\;.
\end{align}
On the contrary, the $\Gamma_3$ operator does not require renormalisation \citep[e.g.][]{assassi2014renormalized} and we have
\begin{align}
P_{\Gamma_3 \delta}(k)=B_{\Gamma_3 \delta}(k)=\lim_{\Lambda\to \infty} P_{\Gamma_3 \delta}^{(31)}(k)\;,
\end{align}
with\footnote{The perturbative contributions to $\Gamma_3$ vanish at second order.}
\begin{align}
P_{\Gamma_3 \delta}^{(31)}(k)=
4P_{11}(k) \int &\left[F_2(-\mathbf{q}, \mathbf{k})-G_2(-\mathbf{q}, \mathbf{k})\right] \nonumber \\ 
&\cdot\left[\left(\frac{\mathbf{q}\cdot(\mathbf{k}-\mathbf{q})}{q\, |\mathbf{k}-\mathbf{q}|}\right)^2-1\right] \, P_{11}(q)\, \frac{\mathrm{d}^3q}{(2 \pi)^3}\;.
\end{align}
A few notes are in order here.
First, all terms proportional to $P_{11}(k)$ have been used to define the renormalised linear bias parameter
as a function of the bare bias coefficients:
\begin{align}
b_1^\mathrm{R}=b_1+\left(\frac{68}{21}b_2+\frac{136}{63} b_{s^2}+3 b_3+\frac{2}{3} b_{\delta s^2} \right)\sigma_1^2(\Lambda) \;.
\label{eq:b1theo}
\end{align}
This makes sure that $P_{\delta_\mathrm{h}\delta}(k)\to b_1^\mathrm{R}\, P_{\delta \delta}(k)$ in the limit $k\to 0$.
Secondly, although the perturbative calculations give $b_1^\mathrm{R}\,P_{\delta \delta}^\mathrm{(LO)}(k)+b_1\,P_{\delta \delta}^\mathrm{(NLO)}(k)$, this expression has been replaced with $b_1^\mathrm{R}\,[P_{\delta \delta}^\mathrm{(LO)}(k)+P_{\delta \delta}^\mathrm{(NLO)}(k)]$ in equation (\ref{eq:fullpert}), the difference being proportional to $\sigma_1^2\,P_{\delta \delta}^\mathrm{(NLO)}(k)$ and thus higher-order in the perturbations.
As we will show later, this substitution is a source of error on mildly non-linear scales.
Thirdly, the third-order bias coefficients $b_3$ and $b_{\delta s^2}$ are renormalised into $b_1^\mathrm{R}$ and do not appear in equation (\ref{eq:fullpert}).
Fourthly, it turns out that the functions $B_{s^2 \delta}(k)$ and $B_{\Gamma_3 \delta}(k)$
are proportional to each other and therefore
a single bias parameter (a linear combination of $b_{s^2}$ and $b_{\Gamma_3}$) could be used to scale their total contribution \citep{mcdonald2009clustering}.
Finally, no direct contributions from the stochastic bias coefficients $\epsilon_{O}$ appear in the r.h.s. of equation (\ref{eq:fullpert}).
This is because these zero-mean stochastic fields are assumed to be independent of $\delta$.
If, however, matter-density perturbations are described in terms of an effective deterministic field $\delta_{\mathrm{eff}}$ plus stochastic contributions $\epsilon_{\mathrm m}$ (from the small-scale modes that are not included in the theory), then 
the correlation between the stochastic terms $\epsilon_1$ and $\epsilon_{\mathrm m}$ should 
contribute to $P_{\delta_\mathrm{h}\delta}^{\mathrm{(nlo)}}(k)$. The corresponding
cross spectrum is represented by the term
$\Upsilon_{2,\epsilon_1,\epsilon_{\mathrm m}}\,k^2=
(\partial \Upsilon_{\epsilon \epsilon_{\mathrm m}}/\partial k^2)_{k=0}\,k^2$ in equation (\ref{eq:fullpert}).
Theoretical considerations suggest
that this term should be
highly suppressed with respect to the other
NLO corrections \citep{senatore2015bias,angulo2015one}.

\subsection{Impact of filter functions}
\label{sec:filters}
Equations (\ref{eq:P6821}) and (\ref{eq:P13663}) are obtained from equations (\ref{eq:P31})
and (\ref{eq:P31s2})
by performing the integration over $\mathbf{q}$ in spherical polar coordinates. The integration range extends over the full solid angle around $\mathbf{k}$ but is limited to the region $q<\Lambda$ for the radial component.
We show here that this calculation does not exactly give what one would obtain by smoothing the density field with a sharp cutoff in $k$ space as, for instance, we will do later when analyzing N-body data.

As mentioned in section \ref{sec:biaseft}, the bias expansion applies to fields that have been low-pass filtered. Let us consider the smoothed field $W(k) \,\delta(\mathbf{k})$ with $W(k)$ a low-pass, spherically symmetric, window function. In this case, equation (\ref{eq:P31}) should be replaced with
\begin{align}
P_{\delta^2 \delta}^{(31)}(k)
=4 P_{11}\big(k\big)W\big(k\big) \int F_2\big(-\mathbf{q}, \mathbf{k}\big) & P_{11}\big(q\big) W\big(q\big) \nonumber \\ & W\big(|\mathbf{k} - \mathbf{q}|\big) \frac{\mathrm{d}^3q}{(2 \pi)^3}\;.
\label{eq:6821sims}
\end{align}
For simplicity, in this work, we use a spherical top-hat filter in $k$ space,
\begin{equation}
W(k)=\begin{cases}
1 &\text{if $k<\Lambda$}\;,\\
0 & \text{otherwise}\;.
\label{eq:sks}
\end{cases}
\end{equation}
The factor $W(|\mathbf{k} - \mathbf{q}|)$ in the integrand of equation~(\ref{eq:6821sims}) thus limits the integration range to the region where
\begin{equation}
\Lambda^2>|\mathbf{k} - \mathbf{q}|^2=k^2+q^2-2qk\mu\; ,
\label{eq:filtereffect}
\end{equation}
with $\mu=(\mathbf{k}\cdot\mathbf{q})/(kq)$ the cosine of the angle between $\mathbf{k}$ and $\mathbf{q}$. This turns out to be a constraint on $\mu$ at fixed $q$, 
\begin{equation}
\mu> \frac{k^2+q^2-\Lambda^2}{2qk}=\mu_\mathrm{min} \;,
\end{equation}
and implies that, for $q>\Lambda-k$, the integration range in $\mu$ will have a lower bound $\mu_\mathrm{min} > -1$. 
The complete result for $0<k<\Lambda$ is
\begin{align}
P_{\delta^2 \delta}^{(31)}(k)
&=P_{11}(k)W(k)\left[
\frac{68}{21}\sigma_1^2(\Lambda-k)\right.\nonumber \\
&\left.+2 \int_{\Lambda-k}^\Lambda \int_{\mu_\mathrm{min}}^1 F_2\big(-\mathbf{q}, \mathbf{k}\big)\,\mathrm{d}\mu\,  P_{11}\big(q\big) \frac{q^2\mathrm{d}q}{2 \pi^2}\right]\;,
\label{eq:expand}
\end{align}
where the integral over $\mu$ reduces to
\begin{align}
\int_{\mu_\mathrm{min}}^1 F_2\big(-\mathbf{q}, \mathbf{k}\big)\,\mathrm{d}\mu&=
\frac{17}{21}-\frac{51}{112} \left(\frac{k}{q}+\frac{q}{k} \right)+\frac{5}{28}\frac{\Lambda^2}{kq}\nonumber\\&+\frac{17}{336}\left(\frac{k^3}{q^3}
+\frac{q^3}{k^3} \right)+\frac{3}{112}\left(\frac{\Lambda^4}{k^3q}+\frac{\Lambda^4}{kq^3} \right)\nonumber\\
&-\frac{5}{56}\left(\frac{\Lambda^2k}{q^3}+\frac{\Lambda^2q}{k^3} \right)+\frac{1}{84}\frac{\Lambda^6}{k^3q^3}\;.
\end{align}
It follows that
\begin{align}
P_{\delta^2 \delta}^{(31)}(k)
&=P_{11}(k)W(k)\left\{
\frac{68}{21}\sigma_1^2(\Lambda-k)\right.\nonumber \\
&+\frac{34}{21} \left[\sigma_1^2(\Lambda)-\sigma_1^2(\Lambda-k) \right] \nonumber \\
&+ \left(\frac{17}{168}k^3-\frac{5}{28}\Lambda^2k+\frac{3}{56}\frac{\Lambda^4}{k}+\frac{1}{42}\frac{\Lambda^6}{k^3} \right)\, \mathcal{H}_{-1}(k)\nonumber \\
&+ \left(-\frac{51}{56}k+\frac{5}{14}\frac{\Lambda^2}{k}+\frac{3}{56}\frac{\Lambda^4}{k^3}\right)\, \mathcal{H}_1(k)  \nonumber \\
& + \left(-\frac{51}{56}\frac{1}{k}-\frac{5}{28}\frac{\Lambda^2}{k^3} \right)\, \mathcal{H}_3(k)\nonumber \\
&\left.+ \frac{17}{168}\frac{1}{k^3}\, \mathcal{H}_5(k)
\right\}\;.
\label{eq:full31}
\end{align}
where
\begin{equation}
\mathcal{H}_n(k)=\int_{\Lambda-k}^\Lambda q^n\, P_{11}(q) \frac{\mathrm{d}q}{2\pi^2}    
\end{equation}
and $\sigma_1^2(\Lambda)-\sigma_1^2(\Lambda-k)=\mathcal{H}_2(k)$. 
By Taylor expanding the $\mathcal{H}_n(k)$ functions to fourth order, we obtain the linear expansion of the expression in the curly parentheses in equation~(\ref{eq:full31}), 
\begin{align}
\label{eq:filter31d2}
\frac{P_{\delta^2 \delta}^{(31)}(k)}
{P_{11}(k)W(k)}
  &=\frac{68}{21}\sigma_1^2(\Lambda)-\frac{1}{3}\frac{\Lambda^3P_{11}(\Lambda)}{2\pi^2}\nonumber\\&+\left[-\frac{17}{28}\frac{\Lambda^2P_{11}(\Lambda)}{2\pi^2}+\frac{1}{8}\frac{\Lambda^3P'_{11}(\Lambda)}{2\pi^2} \right]\,k\\&
+\mathcal{O}(k^2)\;. \nonumber
\end{align}
This result has some implications for the renormalisation of the bias expansion.
First, the resulting expression for $P_{\delta^2 \delta}^{(31)}(k)$ is not proportional to $P_{11}(k)$ at finite wavenumbers. Moreover, the limit for $k\to 0$ of equation (\ref{eq:filter31d2}) is not $(68/21)\, \sigma_1^2(\Lambda)$ 
\citep[contrary to what claimed by][after their equation 2.122]{desjacques2016large}
and a different cutoff-dependent coefficient should be used to renormalise $b_1$. 
Note, however, that, for a $\Lambda$CDM cosmology, the $k\to 0$ limit reduces to (\ref{eq:P6821}) for $\Lambda \to +\infty$. We will return to these issues and present a comparison with N-body simulations in section~\ref{sssec:checkfilter}. Analogous calculations can also be performed for $s^2$ and other composite operators.

\section{Numerical methods}
\label{sec:numerics}
In this section, we introduce our suite of N-body simulations and describe the numerical techniques we use to build the continuous density fields, as well as the auto and cross power spectra that appear in equation~(\ref{eq:fullpert}).

\subsection{N-body simulations}
\label{ssec:sims}

\begin{figure}
  \centering{
      \includegraphics[width=\columnwidth]{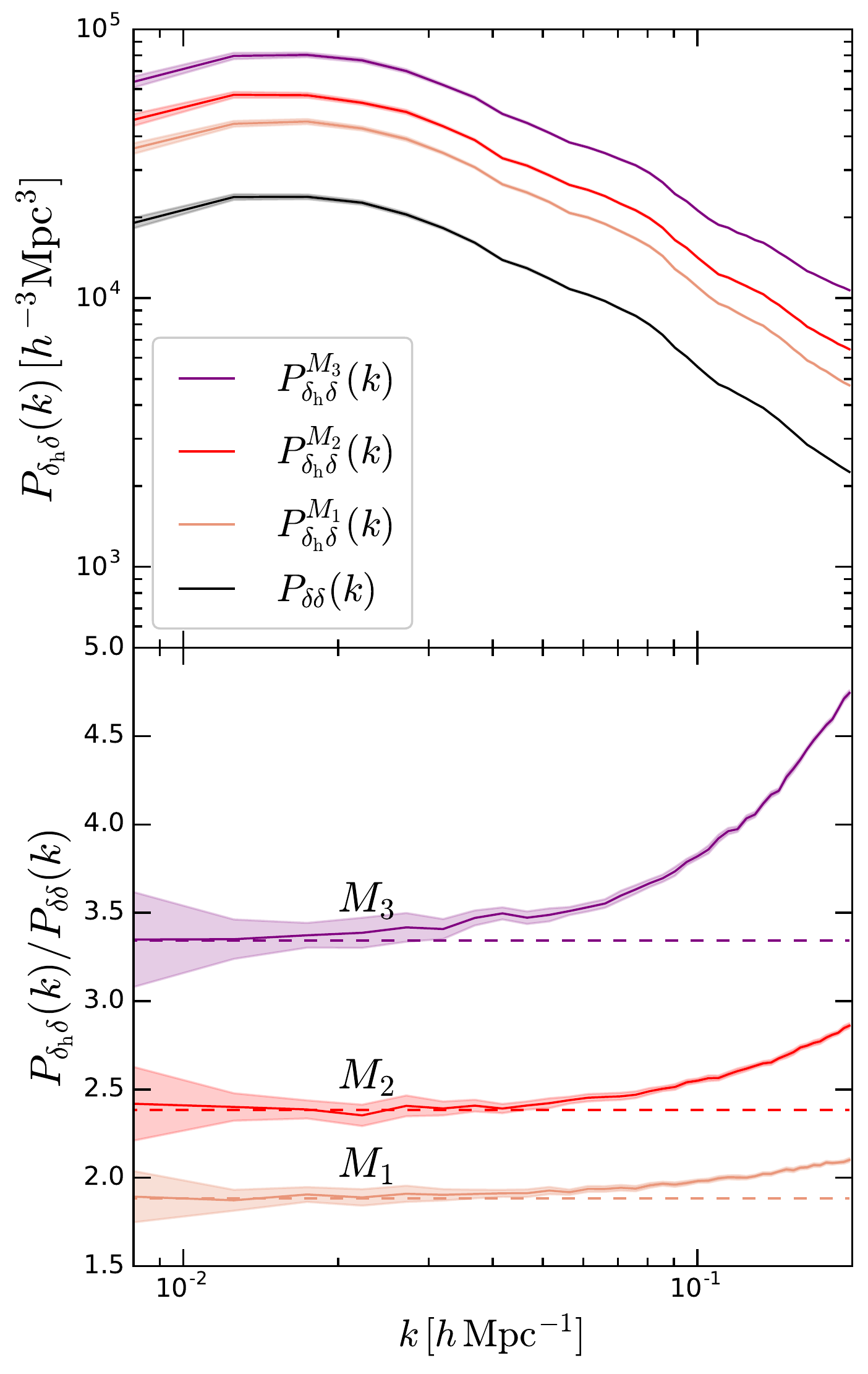}\\
    }
    \caption{Top: Average matter power spectrum $P_{\delta\delta}(k)$ and halo-matter cross spectra $P_{\delta_\mathrm{h}\delta}(k)$ (for the mass bins $M_1, M_2, M_3$  defined in section \ref{ssec:sims}) measured from our 40 simulations. Shaded regions represent the standard error of the mean. Bottom: Corresponding effective bias functions $P_{\delta_\mathrm{h}\delta}(k)/P_{\delta\delta}(k)$. The dashed lines indicate a constant fit to the five leftmost data points.}
    \label{fig:powertransfer}
\end{figure}

\begin{table*}
	\centering
	\caption{The set of parameters that characterise our suite of $N_\mathrm{sim}$ simulations: $L_\mathrm{box}$ denotes the box length, $N_\mathrm{part}$ the number of particles, $M_\mathrm{part}$ the particle mass, $L_\mathrm{soft}$ the softening length, and $z_\mathrm{IC}$ the initial redshift.}
	\label{tab:simdata}
	\begin{tabular}{ccccccc}
		\hline
		$N_\mathrm{sim}$ & $L_\mathrm{box}$ & $N_\mathrm{part}$ & $M_\mathrm{part}$ & $L_\mathrm{soft}$ & $z_\mathrm{IC}$ & Cosmology \\
		\hline
		40 & 1200\mpch & $512^3$ & $1.1037 \times 10^{12} h^{-1}\, \mathrm{M}_\odot$ & 0.078\mpch & 50 & Planck 2015 \\
		\hline
	\end{tabular}
\end{table*}

\begin{figure*}
  \centering{
      \includegraphics[width=0.95\textwidth]{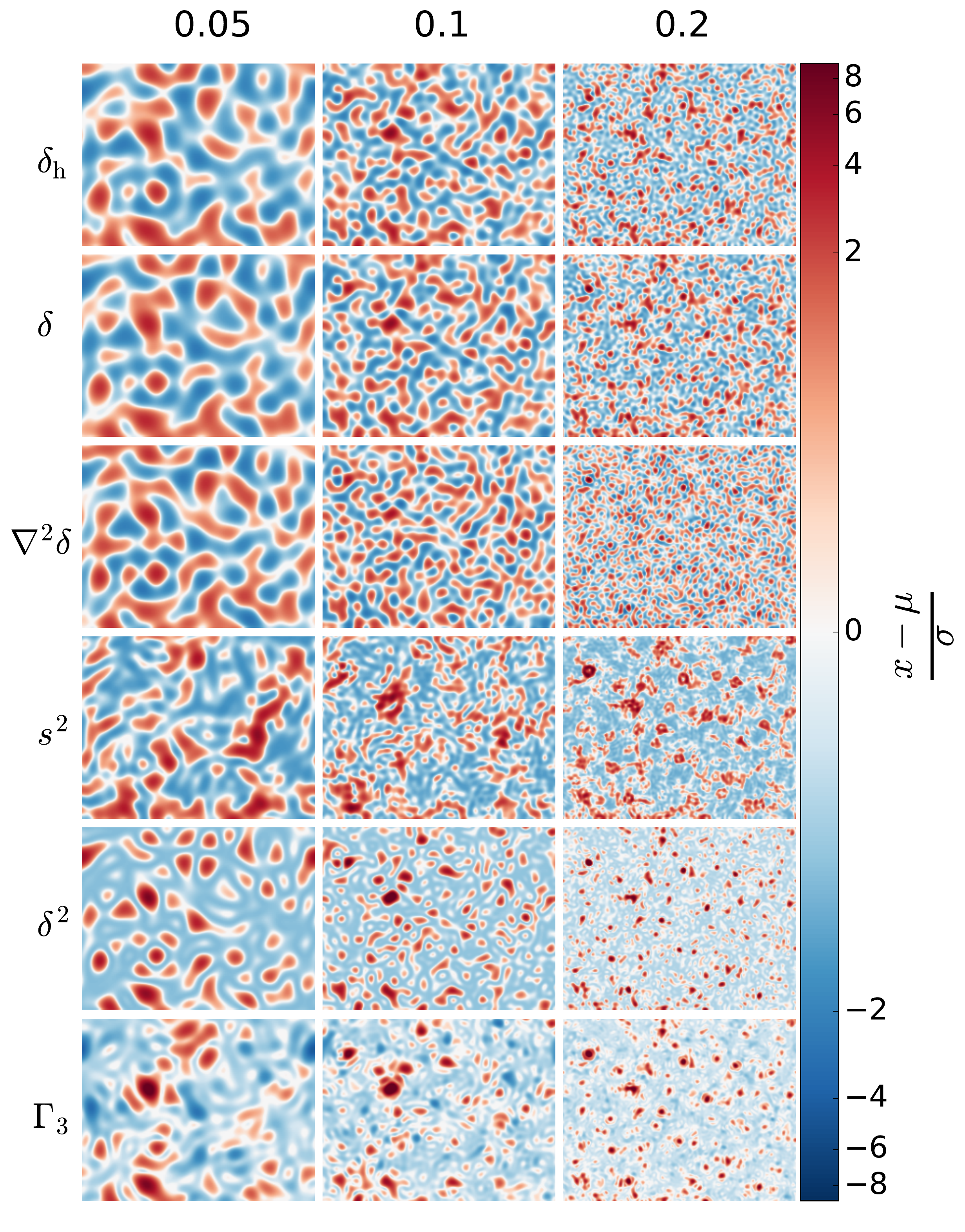}\\
    }
    \caption{The fields that are included in the bias expansion. We show a 2D slice extracted from one N-body simulation by fixing one of the spatial coordinates. The grid of images shows different fields along the vertical direction (as indicated by the labels) and different values of the cutoff scale $\Lambda$ (in \hmpc) in the horizontal direction. All the fields have been standardised to zero mean and unit variance. Note that the images only cover 80 per cent of the box in the vertical direction.}
    \label{fig:simlambda}
\end{figure*}  

We use the SPH code \textsc{gadget-2} (\citealt{springel2005cosmological}) to run 40 DM-only simulations. Their main characteristics are summarised in Table~\ref{tab:simdata}. We set up the initial conditions (IC) at $z_\mathrm{IC}=50$ with the \textsc{music} code (\citealt{hahn2011multi}) employing the Planck 2015 (\citealt{ade2016planck}) cosmology, i.e. $h=0.677$, $\sigma_8=0.816$, $n_s=0.967$, $\Omega_\mathrm{m}=0.3089$, $\Omega_\mathrm{b}=0.0486$ and $\Omega_\Lambda=0.6911$. 

To identify gravitationally bound structures, we use the \textsc{rockstar} halo finder \citep{behroozi2013rockstar}. This algorithm uses the phase-space distribution of the simulation particles to detect spherical haloes with a virial mass defined as in \citet{bryan1998statistical}. We split the halo population into three mass bins based on the number of halo particles. In the lowest-mass bin, $M_1$, haloes contain between 40 and 79 particles, corresponding to a mean mass of $\overline{M}_1= 6.15\times 10^{13} h^{-1} M_\odot$. The second bin includes haloes with 80 to 159 particles ($\overline{M}_2=1.21\times 10^{14} h^{-1} M_\odot$), while more massive structures with $\geq 160$ particles fall into the highest-mass bin ($\overline{M}_3=3.17\times 10^{14} h^{-1} M_\odot$).

\subsection{Measuring smoothed fields and spectra}
\label{ssec:fieldspectra}

\begin{table}
	\centering
	\caption{The variance of the matter density field as a function of the cutoff scale $\Lambda$ at $z=0$. We contrast the results obtained from linear SPT ($\sigma^2_1$) with those measured in the simulations ($\sigma^2_\mathrm{sim}$).}
	\begin{tabular}{ccc}
	\hline
	$\Lambda\, [h \, \mathrm{Mpc}^{-1}]$ & $\sigma_1^2$ & $\sigma_{\mathrm{sim}}^2$ \\
	\hline 
	0.05 & 0.0377 & 0.0343 \\
	0.10 & 0.1672 & 0.1513 \\
	0.20 & 0.5430 & 0.5223 \\
	\hline
	\end{tabular}
	\label{tab:sigma2}
\end{table}

We employ a `cloud-in-cell' (CIC) interpolation method to build the gridded overdensity fields $\delta$ and $\delta_{\mathrm h}$ starting from the positions of the N-body particles and of the haloes, respectively. The fields are sampled on a regular Cartesian mesh with $256^3$ cells that fully covers the simulation box. After correcting for the mass-assignment scheme, we use the fast-Fourier-transform (FFT) algorithm to obtain $\delta(\mathbf{k})$ and $\delta_{\mathrm h}(\mathbf{k})$. In the top panel of Fig.~\ref{fig:powertransfer}, we show the matter power spectrum $P_{\delta\delta}(k)$ and the halo-matter cross spectra $P_{\delta_\mathrm{h}\delta}(k)$ averaged over our set of simulations for each halo mass bin. The shaded regions indicate the standard error of the mean. In the bottom panel, we show the ratio $P_{\delta_\mathrm{h}\delta}(k)/P_{\delta\delta}(k)$ which can be interpreted as an effective scale-dependent bias. This function grows with $k$, and this effect becomes more prominent for more massive haloes. In order to emphasize this trend, we fit a constant to the five leftmost data points and plot the result with a dashed line. The discrepancy between the solid and dashed line provides a strong motivation for considering non-linear bias models. 

Low-pass smoothing is applied in Fourier space by multiplying the FFT of the fields by the window function $W(k)$ given in equation (\ref{eq:sks}). We use three different values for the cutoff scale $\Lambda$ ensuring that the variance of the $\delta$ field is smaller than unity (see Table~\ref{tab:sigma2}). We also apply spectral methods to compute the Fourier transforms of $s_{ij}$ and $\nabla^2 \delta$ starting from $\delta(\mathbf{k})$. In order to compute $\Gamma_3$, we first build three momentum grids (one for each Cartesian component of the momentum vector) by applying the CIC interpolation to the particle velocities. To get the velocity grids from these we divide the momentum components by the density. We apply the FFT to these grids so that we can first compute $\theta(\mathbf{k}) = i\mathbf{k} \cdot \mathbf{v}(\mathbf{k})/(aHf)$ and then $p_{ij}(\mathbf{k})$. Finally, we transform all the smoothed fields back to real space and compute the quadratic fields $\delta^2(\mathbf x)$, $s^2(\mathbf x)$, $p^2(\mathbf{x})$ and $\Gamma_3(\mathbf{x})$.

A sample slice extracted from one of the simulation boxes at $z=0$ for all fields is shown in Fig.~\ref{fig:simlambda}. From top to bottom we plot the different fields, from left to right we change the cutoff scale $\Lambda$. We standardised the fields such that dark tones indicate large deviations from than the mean. As expected, larger values of $\Lambda$ give rise to more detailed structure in all panels. The most striking feature is that $\delta_\mathrm{h}$ and $\delta$ always look very similar. This is not surprising since $\delta_\mathrm{h} \propto \delta $ at LO in the bias expansion. For larger $\Lambda$, the similarity is less evident, reflecting the increased importance of higher-order and derivative terms. The Laplacian of $\delta$ looks qualitatively similar to $\delta$ for small $\Lambda$, whereas, for large $\Lambda$, short-wavelength Fourier modes are enhanced with respect to the density field. The field $\delta^2$ presents concentrated high positive peaks on top of a rather uniform background, somewhat reminiscent of shot noise \citep[see also][]{heavens1998non}. Around those peaks the signal of $s^2$ becomes strong in almost spherical shells which are particularly evident for large $\Lambda$. Finally, for $\Lambda=0.2$\hmpc, $\Gamma_3$ presents high peaks at the same locations as $\delta^2$ and $s^2$ while this correspondence is less striking for smaller $\Lambda$. Not visible in the figure is that the relative amplitudes of the different fields change dramatically with the cutoff scale. Whereas, for $\Lambda=0.05$\hmpc, they are clearly ordered as expected from SPT, for $\Lambda=0.2$\hmpc, $\delta$, $\delta^2$, $s^2$ and $\Gamma_3$ are all of order unity.

We employ standard methods to compute the auto and cross power spectra between the fields using 10, 20 and 40 linearly-spaced bins in $k$ for $\Lambda = 0.05,\, 0.1,$ and $0.2$\hmpc, respectively.

\section{Renormalisation in simulations}
\label{sec:simrenorm}

\begin{figure*}
  \centering{
      \includegraphics[width=\textwidth]{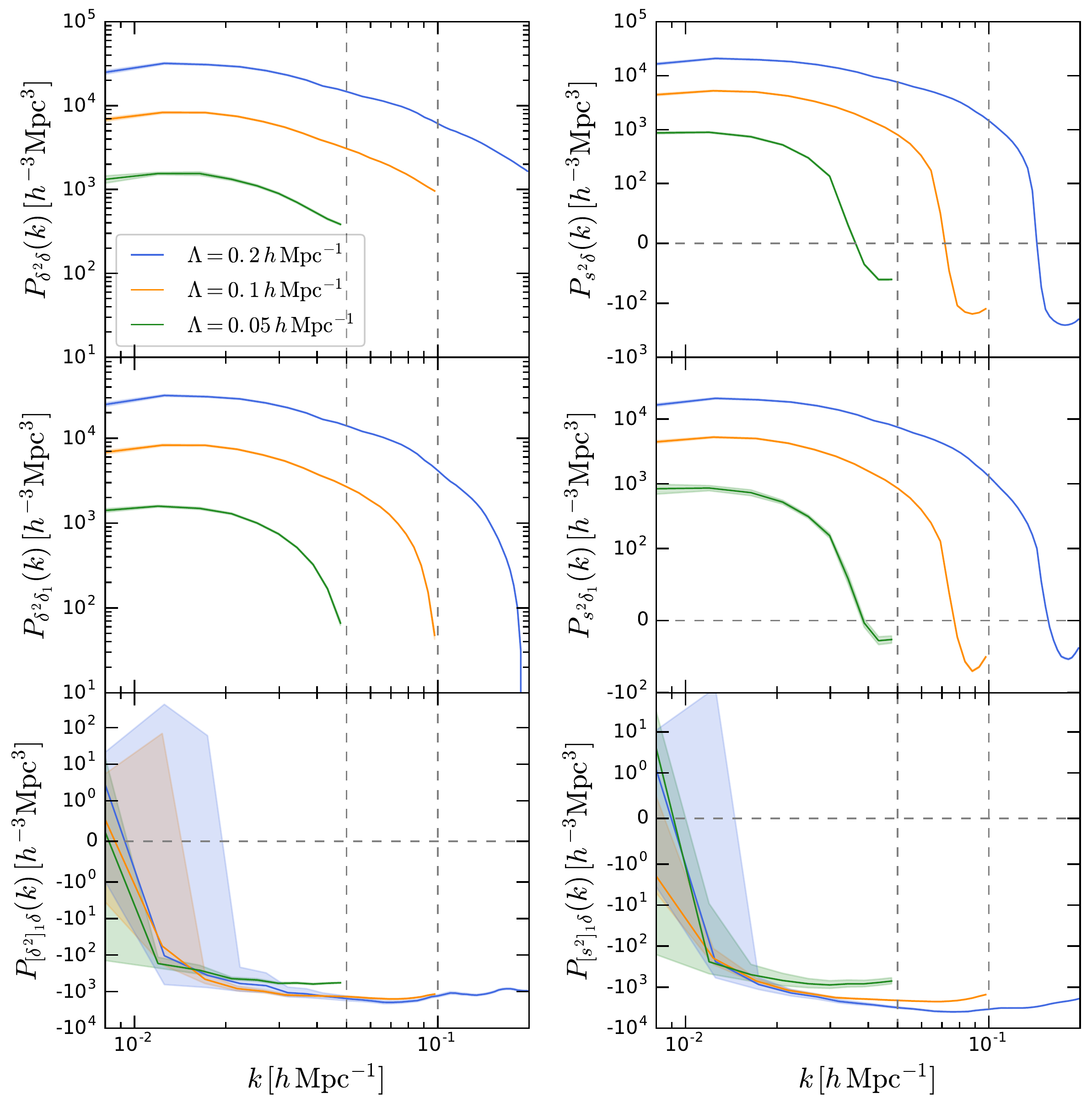}\\
    }
    \caption{Average cross spectra $P_{\delta^2 \delta}(k)$ and $P_{s^2 \delta}(k)$, measured from our 40 simulations for different values of $\Lambda$. Shaded regions represent the standard error of the mean. Top: The `full' spectra, obtained from cross-correlating the quadratic operators $\delta^2$ and $s^2$ with $\delta$. Middle: The UV-sensitive part of the spectra, obtained by correlating $\delta^2$ and $s^2$ with $\delta_1$. Bottom: The renormalised spectra, obtained by subtracting the UV-sensitive terms from the full spectra. A symmetric logarithmic scale on the $y$ axis is used for quantities that assume both positive and negative values. The vertical dashed lines indicate the locations of the different cutoff scales.}
    \label{fig:spectraall}
\end{figure*}

In this section, we measure the UV-sensitivity of the composite operators and implement the renormalisation framework directly in our simulations.

In the top panels of Fig.~\ref{fig:spectraall}, we show the average over the 40 simulations of the cross spectra between different composite operators and $\delta$ for three values of $\Lambda$. We plot $P_{\delta^2 \delta}(k)$ on the left-hand side and $P_{s^2 \delta}(k)$ on the right-hand side. As expected, our results are UV-sensitive as the amplitude of the spectra changes dramatically with $\Lambda$. Qualitatively, this is consistent with the perturbative calculations presented in section~\ref{sec:COSPT}.

We want to build a procedure that removes the UV-sensitive part from the cross spectra. In the perturbative calculations, this term is
$[\lim_{q\to 0}
P_{{O} \delta}^{(31)}(q)/P_{11}(q)]\,P_{11}(k)$.
For instance, if  ${O}=\delta^2$, then $P_{{O} \delta}^{(31)}(q)$ is obtained from the correlator $2 \langle(\delta_2 \delta_1)\, \delta_1 \rangle$ (parentheses here denote the fields contributing to $\delta^2$). 
This is the LO term of $\langle \delta^2 \delta_1 \rangle$. 
In the simulations, we cannot isolate $\delta_2$ from all the other non-linear terms. However, if the perturbative ansatz holds true on the largest scales, we can safely assume that $\langle \delta^2 \delta_1 \rangle$ is dominated by the LO part when $k\to 0$. Therefore, it makes sense to measure $P_{\delta^2 \delta_1}(k)$ from the simulations and compute the `renormalised cross spectrum'
\begin{align}
    P_{[\delta^2]_1 \delta}(k) &= P_{\delta^2 \delta}(k) - \frac{P_{\delta^2 \delta_1}(k_\mathrm{min})}{P_{11}(k_\mathrm{min})}P_{11}(k)\nonumber \\&= P_{\delta^2 \delta}(k) - \alpha_{\delta^2}(\Lambda)\,P_{11}(k)\;,
    \label{eq:renormd2}
\end{align}
where $k_\mathrm{min}$ denotes the bin containing the lowest wavenumbers that can be accessed in our simulation boxes and $P_{11}(k)$ is computed directly by re-scaling the IC of the simulations. 
Similarly, for $s^2$, we have 
\begin{align}
P_{[s^2]_1 \delta}(k) &= P_{s^2 \delta}(k) - \frac{P_{s^2 \delta_1}(k_\mathrm{min})}{P_{11}(k_\mathrm{min})}P_{11}(k)\nonumber \\
&= P_{s^2 \delta}(k) - \alpha_{s^2}(\Lambda)\,P_{11}(k)
\;.
\label{eq:renorms2}
\end{align}

\subsection{Measuring the linear growth factor}
\label{sec:growth}

\begin{figure}
  \centering{
      \includegraphics[width=\columnwidth]{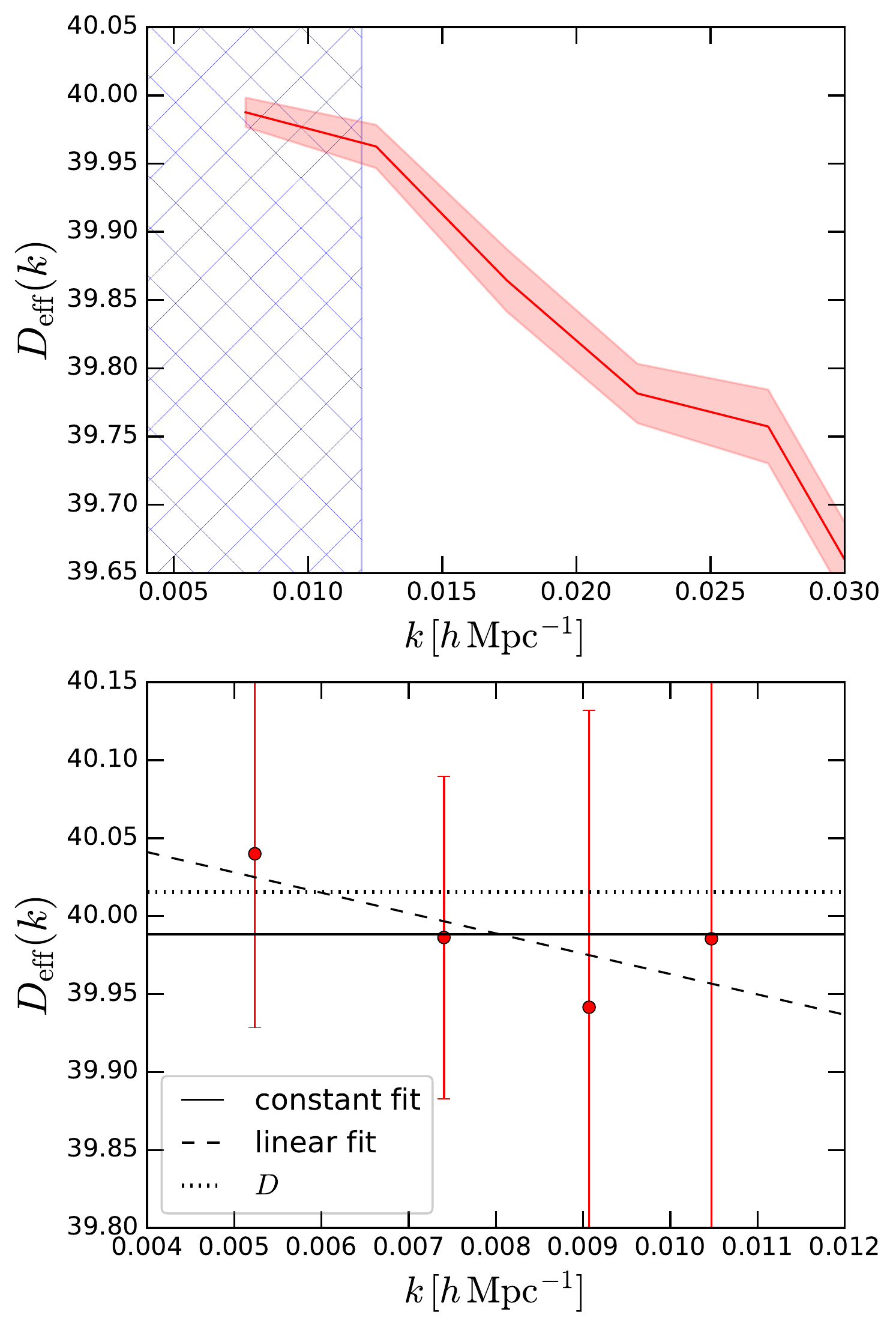}\\
    }
    \caption{The top panel shows the $k$-dependent growth factor $D_\mathrm{eff}(k)$, measured in our simulations by using equation~(\ref{eq:Dk}). The solid line and the shaded region indicate the average over the 40 realisations and its error, respectively. The hatched region is magnified in the bottom panel where we fit a constant (solid) and a linear function of $k$ (dashed) to  $D_\mathrm{eff}(k)$
    measured at the first four multiples of the fundamental frequency $2\pi/L_\mathrm{box}$ (data points with error bars). The dotted line represents the linear growth factor for the adopted cosmology, $D=40.015$.}
    \label{fig:growth}
\end{figure}

In order to implement the renormalisation procedure sketched above, we need the linear density field $\delta_1$ evaluated at $z=0$ in the N-body simulations. This is obtained by multiplying the IC by the linear growth factor $D$. Although $D$ can be computed from the cosmological parameters by performing a numerical integration ($D=40.015$ at $z=0$), the result does not necessarily match the actual growth of perturbations in the simulations which only provide an approximate solution. Previous studies have shown that different N-body codes might differ by up to 0.3 per cent in the large-scale normalization of the matter power spectrum at $z=0$ which corresponds to a 5 per cent error on $D$ \citep[see e.g. Fig.~1 in][]{schneider2016matter}. Since the success of the renormalisation procedure relies on the subtraction of two signals with comparable amplitudes to get a much smaller one, we need to make sure that we know the precise value of the growth factor realized in the simulations. Therefore, we proceed as follows. First, we measure an `effective' scale-dependent growth function directly from our simulations as
\begin{equation}
    D_{\mathrm{eff}}(k)=\Big\langle\frac{\delta(\mathbf{k},z=0)}{\delta_1(\mathbf{k},z_\mathrm{IC})}\Big\rangle_{k\,\mathrm{bin}}
    \label{eq:Dk}
\end{equation}
where the average is taken over the same bins of $k$ as the spectra and we have explicitly written the redshift at which the fields are evaluated. The result for $k<0.03\,h$ Mpc$^{-1}$ is shown in the top panel of Fig.~\ref{fig:growth} where the shaded region indicates the error of the mean over the simulations. Note that, as expected from theoretical considerations,
$D_\mathrm{eff}$ decreases with $k$ on these scales. We are interested in the limit of this function for $k\to 0$. For this reason, we only consider the leftmost bin and analyze the signal at the level of single Fourier modes (see the bottom panel of Fig.~\ref{fig:growth}). In order to extract the limit, we fit a constant ($D_{\mathrm{eff}}=39.98\pm 0.07$, dotted) and a linear function of $k$ ($D_\mathrm{eff}(k)=40.09 \pm 0.31 - (13 \pm 39)\, (k/1\,h\, \mathrm{Mpc}^{-1})$, dashed) to the measurements.
The slope of the linear fit is consistent with zero within the errors. Furthermore, the two fits give consistent limits for $k\to 0$ which are also compatible with the expected value for $D$.
For our calculations, we therefore use $D=39.98$.

\subsection{Renormalising the spectra}
\label{sssec:renormspectra}

We are now ready to compute $P_{\delta^2 \delta_1}(k)$ and $P_{s^2 \delta_1}(k)$ as well as the renormalised spectra $P_{[\delta^2]_1 \delta}(k)$ and $P_{[s^2]_1 \delta}(k)$.
A note is in order here. The IC of our simulations are Gaussian and all the three-point correlators of $\delta_1$ should, in principle, vanish. In practice, however, we measure very noisy non-zero values for $P_{\delta_1^2 \delta_1}(k)$ and $P_{s_1^2 \delta_1}(k)$. In the perturbative framework, these spectra coincide with the tree-level terms of $P_{\delta^2 \delta}(k)$ and $P_{s^2 \delta}(k)$. Therefore, in order to improve the quality of our results, we subtract the noisy terms from all the non-linear cross spectra appearing on the r.h.s. of equations~(\ref{eq:renormd2}) and (\ref{eq:renorms2}). Our final results for $P_{\delta^2 \delta_1}(k)$ and $P_{s^2 \delta_1}(k)$ are shown in the middle panels of Fig.~\ref{fig:spectraall}. As expected, they show a similar behaviour as a function of both $\Lambda$ and $k$ with respect to $P_{\delta^2 \delta}(k)$ and $P_{s^2 \delta}(k)$. However, when $k$ approaches $\Lambda$ their signal is suppressed.
The corresponding values of $\alpha_{\delta^2}(\Lambda)$ and $\alpha_{s^2}(\Lambda)$ are presented in Table~\ref{tab:alpha}.

\begin{table}
	\centering
	\caption{The factors $\alpha_{\delta^2}(\Lambda)$ and $\alpha_{s^2}(\Lambda)$ defined in equations~(\ref{eq:renormd2}) and (\ref{eq:renorms2}). The values indicate the average over the simulations and the uncertainty is the standard error of the mean.}
	\begin{tabular}{ccc}
	\hline
	$\Lambda\, [h \, \mathrm{Mpc}^{-1}]$ & $\alpha_{\delta^2}(\Lambda)$ & $\alpha_{s^2}(\Lambda)$ \\
	\hline 
	0.05 & $0.072\pm 0.001$ & $0.046\pm 0.001$  \\
	0.10 & $0.349\pm 0.006$ & $0.228\pm 0.004$ \\
	0.20 & $1.261\pm 0.017$ & $0.834\pm 0.013$\\
	\hline
	\end{tabular}
	\label{tab:alpha}
\end{table}

We finally compute the renormalised spectra $P_{[\delta^2]_1 \delta}(k)$ and $P_{[s^2]_1 \delta}(k)$ from equations~(\ref{eq:renormd2}) and (\ref{eq:renorms2}). We first obtain them for each simulation individually and then perform an average. The results are shown in the bottom panels of Fig.~\ref{fig:spectraall}. At large scales, spectra obtained with different values of $\Lambda$ are now compatible within the errorbars while they differ when $k\simeq \Lambda$. This means that the renormalisation procedure was successful.

\subsection{Comparing spectra from simulations and SPT}
\label{sssec:checkfilter}

\begin{figure*}
    \centering{
        \includegraphics[width=\textwidth]{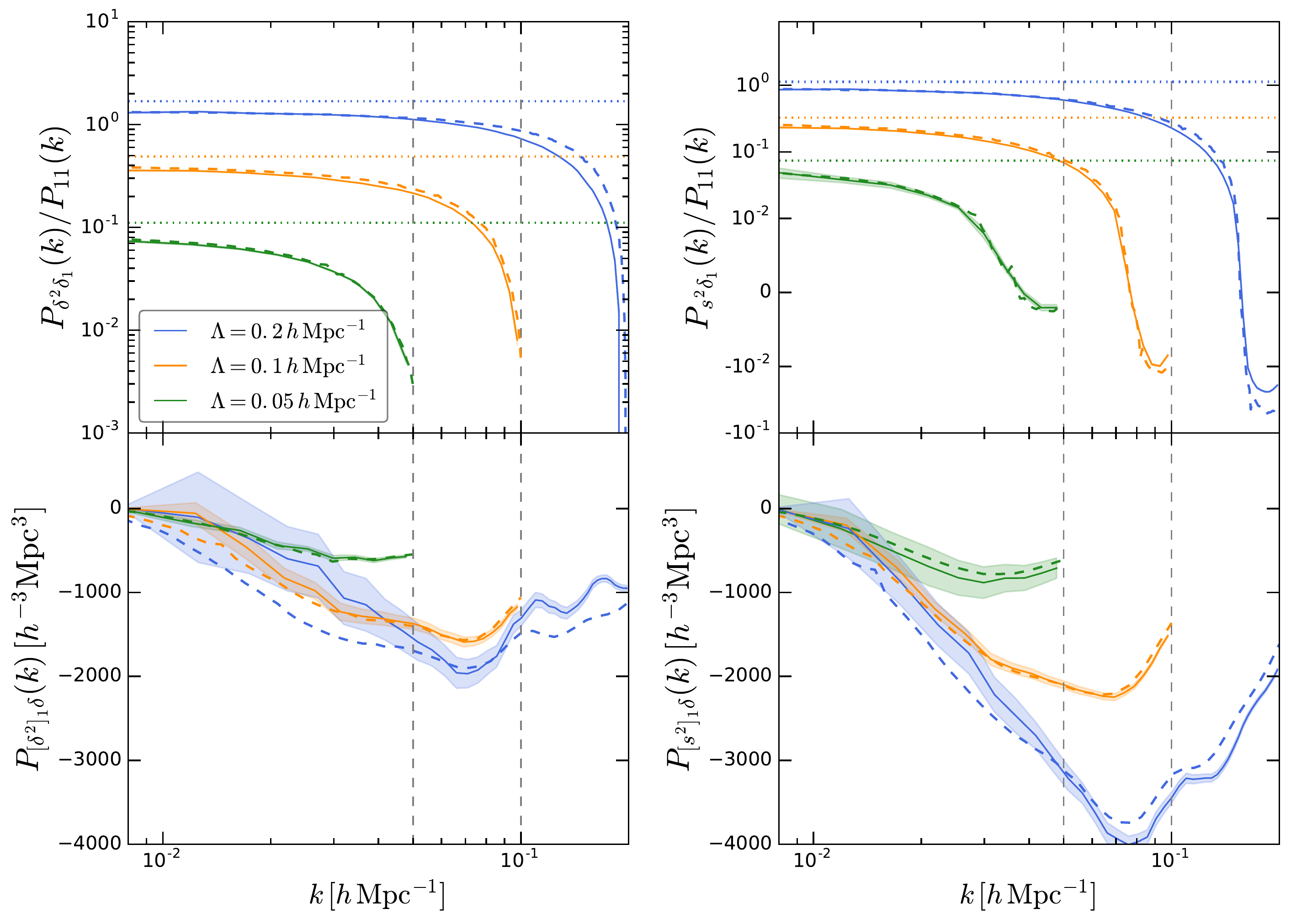}\\
      }
      \caption{The top panels show the UV-sensitive spectra $P_\mathrm{\delta^2 \delta_1}(k)$ (left) and $P_\mathrm{s^2 \delta_1}(k)$ (right) divided by the linear matter power spectrum. Plotted are the average spectra measured from the simulations for different values of $\Lambda$ (solid). Shaded regions represent the standard error of the mean. For comparison, the dashed curves represent $P^{(31)}_{\delta^2 \delta}(k)/P_{11}(k)$ and $P^{(31)}_{s^2 \delta}(k)/P_{11}(k)$ computed from SPT accounting for the filter functions -- see equation~(\ref{eq:6821sims}) and the analogue for $s^2$. Likewise, the horizontal dotted lines indicate the values $68/21\, \sigma^2$ and $136/63\,\sigma^2$ -- see equations~(\ref{eq:P6821}) and (\ref{eq:P13663}). Differently, the vertical dashed lines mark the locations of the different cutoff scales.
      The bottom panels show the renormalised spectra $P_{[\delta^2]_1 \delta}(k)$ (left) and $P_{[s^2]_1 \delta}(k)$ (right) measured from the simulations (solid). Shaded regions represent the standard error of the mean. Also plotted are the perturbative results evaluated by taking the filter functions into account (dashed).}
      \label{fig:renormtest}
\end{figure*} 

In the top-left panel of Fig.~\ref{fig:renormtest}, we show a comparison between $P^{(31)}_{\delta^2 \delta}(k)$, calculated with SPT (equation~(\ref{eq:P6821}), dashed-dotted), with SPT including filter functions (equation~(\ref{eq:6821sims}), dashed), and $P_{\delta^2 \delta_1}(k)$ that we measure from the simulations (solid). We repeat the analysis in the {top right} panel for $P^{(31)}_{s^2 \delta}(k)$ and $P_{s^2 \delta_1}(k)$. Equations~(\ref{eq:P6821}) and (\ref{eq:P13663}) show that in SPT the UV-sensitive terms can be written as a $\Lambda$-dependent prefactor times $P_{11}(k)$. Therefore, in order to conveniently compare the numerical results to the SPT predictions, we always plot the ratios of the cross spectra with $P_{11}(k)$. The measurements from the simulations (solid) decrease with $k$ and are in excellent agreement with the SPT results when filter functions are taken into account (dashed). Note that when $k\to 0$, the curves do not assume the values $68/21\, \sigma^2$ and $136/63\, \sigma^2$ (dotted), but stay below them (as described by e.g. the first line of equation~(\ref{eq:filter31d2}) for $\delta^2$). The logarithmic difference between the solid and dotted curves decreases for larger values of $\Lambda$. This result confirms the discussion presented in section~\ref{sec:filters}.

In the bottom panels of Fig.~\ref{fig:renormtest} we compare the renormalised spectra measured from the simulations (shown already in the bottom panels of Fig.~\ref{fig:spectraall} in logarithmic scale) with the theoretical predictions including filter functions. We find good agreement at all $k$ for small values of $\Lambda$. For $\Lambda=0.2$\hmpc, however, the theoretical results differ more markedly from the simulation measurements. This suggests either that higher-order terms should be accounted for in order to match the N-body results, or that the perturbative expansion starts breaking down at these scales and redshifts.

\section{Measuring bias parameters}
\label{sec:measureb}

We now measure the bias parameters by fitting $P_{\delta_\mathrm{h}\delta}(k)$ with different bias models, renormalised and not. First, we describe our Bayesian fitting routine that takes into account the correlations between the different spectra extracted from our simulations. Subsequently, we discuss how many bias parameters are needed to optimally describe our data by using Bayesian model-comparison techniques. Finally, we present the best-fitting parameters.

\subsection{Fitting method}
\label{ssec:fitting}

In this section, we introduce the Bayesian method we use to fit the bias parameters. Readers who are more interested in the results than in the statistical techniques can directly move to section~\ref{ssec:barebias}.

Our goal is to fit a model for $P_{\delta_\mathrm{h}\delta}(k)$ to our simulation suite, that can schematically be written as $P_{\delta_\mathrm{h}\delta}(k) = \sum_O b_O\, P_{O\delta}(k)$. Contrary to previous work in the field, we do not rely on perturbation theory, but we measure the spectra $P_{O\delta}(k)$ together with $P_{\delta_\mathrm{h}\delta}(k)$ from the simulation boxes. This means that both the dependent and the independent variables in our fit are affected by statistical errors. Since these measurement errors are correlated (as they originate from the same underlying density field), we need to account for their covariance matrix \mathbfss{C}. In practice, we treat each realisation as a repeated measure of all the spectra and we infer the best-fitting parameters by writing a global likelihood function for the bias parameters. 

To simplify the notation, we write all the cross spectra for a given $k$ bin as a multi-dimensional vector $\mathbfit{y}$ and their noisy estimates as $\hat{\mathbfit{y}}=\mathbfit{y}+\boldsymbol{\eta}$, where $\boldsymbol{\eta}$ denotes the measurement error. We assume that the estimator of all $y_i$ is unbiased and that the errors are drawn from a multi-variate Gaussian distribution with covariance matrix \mathbfss{C} as stated above. It follows that the probability of measuring $\hat{\mathbfit{y}}$ is
\begin{equation}
\mathcal{P}(\hat{\mathbfit{y}})=\frac{1}{4\pi^2 \sqrt{\mathrm{det} \, \mathbfss{C}}} \exp \Bigg[-\frac{1}{2}(\hat{\mathbfit{y}}-\mathbfit{y})^\mathrm{T} \mathbfss{C}^{-1} (\hat{\mathbfit{y}}-\mathbfit{y}) \Bigg].
\end{equation}
We now order the components of the vectors so that $P_{\delta_\mathrm{h}\delta}(k)$ appears first. We can thus write $y_1=\beta_2 y_2 + \beta_3 y_3 + \dots+\beta_{n} y_n$ (where the coefficient $\beta_i$ represent
the $n-1$ bias parameters that characterize a model). The likelihood function for the model parameters ($\boldsymbol{\beta}, \mathbfit{y}$) given the data is then
\begin{equation}
\mathcal{L}(\boldsymbol{\beta}, \mathbfit{y}) \propto \mathcal{P}(\hat{\mathbfit{y}} | \boldsymbol{\beta}, \mathbfit{y}) = \frac{1}{4\pi^2 \sqrt{\mathrm{det} \, \mathbfss{C}}} \exp \Bigg[-\frac{A}{2}\Bigg] \, ,
\end{equation}
with\footnote{From now on Greek indices run from 1 to $n$ and Roman indices run from 2 to $n$. Moreover, $C_{ij}^{-1}$ denotes the $ij$-element of the inverse of \mathbfss{C}.}
\begin{align}
A = & \, C_{11}^{-1}(\hat{y}_1-\beta_i y_i)^2 + 2 \, C_{1j}^{-1}(\hat{y}_1-\beta_i y_i)(\hat{y}_j- y_j) \nonumber \\
&+ C_{ij}^{-1} (\hat{y}_i- y_i)(\hat{y}_j- y_j) \, .
\end{align}
We then marginalize over the unknown true spectra $\mathbfit{y}$, i.e.
\begin{equation}
\mathcal{L}(\boldsymbol{\beta}) \propto \int_{-\infty}^\infty \mathcal{L}(\boldsymbol{\beta}, \mathbfit{y})\, \mathrm{d}^n y \, .
\label{eq:Lint}
\end{equation}
This is easily achieved after rewriting the integrand in the form of a $n$-dimensional Gaussian distribution. We first rewrite $A$ as
\begin{equation}
A=\hat{y}_\mu C_{\mu \nu}^{-1} \hat{y}_\nu + y_i Q_{ij} y_j -2 \omega_i y_i\; ,
\label{eq:A}
\end{equation}
where 
\begin{equation}
Q_{ij}=C_{ij}^{-1} + C_{11}^{-1} \beta_i \beta_j + \beta_i C_{1j}^{-1} + \beta_j C_{1i}^{-1}
\end{equation}
and
\begin{equation}
\omega_i=C_{i\mu}^{-1} \hat{y}_\mu + \beta_i C_{1\mu}^{-1} \hat{y}_\mu \, .
\end{equation}
We then express the second term on the r.h.s. of equation~(\ref{eq:A}) as a function of a generic vector $\mathbfit{d}$ such that
\begin{equation}
y_i Q_{ij} y_j = (y_i-d_i) Q_{ij} (y_j-d_j) + d_i Q_{ij} y_j + y_i Q_{ij} d_j - d_i Q_{ij} d_j
\end{equation}
and impose that $d_i Q_{ij} y_j + y_i Q_{ij} d_j = 2 \omega_i x_i$. This gives $d_i(Q_{ij}+Q_{ji})y_j = 2 d_iQ_{ij}y_j = 2\omega_i y_i$, implying that $d_i Q_{ij}=\omega_j$. We finally obtain
\begin{equation}
d_i=\omega_j Q_{ji}^{-1} \;,
\end{equation}
which, inserted into equation~(\ref{eq:A}), gives
\begin{equation}
A=\hat{y}_\mu C_{\mu \nu}^{-1} \hat{y}_\nu +(y_i - d_i)Q_{ij}(y_j-d_j) - d_i Q_{ij} d_j \, .
\end{equation}
We now perform the integration in equation~(\ref{eq:Lint}) and arrive at the final expression for one $k$ bin:
\begin{equation}
\mathcal{L}(\boldsymbol{\beta}) \propto {[2\pi \, \mathrm{det} \, (\mathbfss{C} \mathbfss{Q})]}^{-1/2} \, \exp \Bigg[ -\frac{1}{2} \Big(\hat{y}_\mu C_{\mu \nu}^{-1} \hat{y}_\nu - d_i Q_{ij} d_j\Big) \Bigg] \; .
\end{equation}
To combine all $k$ bins, we consider the total likelihood $\mathscr{L}$ defined as
\begin{equation}
\mathscr{L}(\boldsymbol{\beta}) = \prod_{j=1}^{N} \mathcal{L}_j(\boldsymbol{\beta})
\end{equation}
where $N$ denotes the number of $k$ bins. In practice, we replace the unknown covariance matrix $\mathbfss{C}$ with an unbiased estimate derived from the 40 simulations.

We sample the posterior distribution of the bias parameters by using our own Markov Chain Monte Carlo (MCMC) code and assuming flat priors.

\subsection{How many bias parameters are needed?}
\label{ssec:testg3k2}

\begin{table*}
    	\centering
    	\caption{The difference $\Delta$WAIC between a model and the preferred one (highlighted with a dash) obtained by fitting $P_{\delta_\mathrm{h} \delta}$ for mass bin $M_3$. Columns two to five indicate which operators are included in each model. Columns six to eight (as well as nine to eleven) refer to different values of $\Lambda$ expressed in units of \hmpc.}
    	\label{tab:waic}
    	\begin{tabular}{lccccrcccccc}
    	   \hline
    	   Model & $\delta$ & $\delta^2$ & $s^2$ & $\nabla^2 \delta$ & \multicolumn{3}{c}{$\Delta$WAIC (NR \& RNL)} & \multicolumn{3}{c}{$\Delta$WAIC (RL)} \\
    	   \hline
           &&&&& $\Lambda=$ 0.05 & 0.1 & 0.2 & 0.05 & 0.1 & 0.2 \\
    	   \hline
    	   \hline
           $\mathcal{M}_1$ & x & & & & 62.56 & 1672.15 & 69339.13 &                 34.99 & 1683.11 & 69847.85 \\
    	   \hline
    	   
           $\mathcal{M}_{2\mathrm{a}}$ & x & x & & & 83.16 & 492.70 & 4302.72 &     4.53 & 689.03 & 5188.85 \\    	  
    	   \hline
           $\mathcal{M}_{2\mathrm{b}}$ & x & & x & & 40.13 & 449.46 & 9308.73 &     32.54 & 820.53 & 7367.62 \\ 
    	   \hline
           $\mathcal{M}_{2\mathrm{c}}$ & x & & & x & 23.79 & 81.89 & 846.68 &       1.03 & -- & 576.55 \\ 
    	   \hline      	   
    	   
           $\mathcal{M}_{3\mathrm{a}}$ & x & x & x && 12.94 & 436.73 & 1797.75 &    7.24 & 450.06 & 4406.97 \\ 
    	   \hline    	   
           $\mathcal{M}_{3\mathrm{b}}$ & x & x & & x & 10.42 & 40.97 & 828.06 &     -- & 1.63 & 575.36 \\ 
    	   \hline    	   
           $\mathcal{M}_{3\mathrm{c}}$ & x & & x & x & 25.14 & 82.82 & 328.83 &     2.33 & 1.91 & 161.92 \\    	   
    	   \hline
    	   
           $\mathcal{M}_{4}$ & x & x & x & x & -- & -- & -- &                       1.59 & 3.86 & -- \\ 
    	   \hline
    	\end{tabular}
\end{table*}

\begin{table}
    	\centering
    	\caption{As in Table~\ref{tab:waic}, but considering models that include more operators (see columns two and three) with respect to  $\mathcal{M}_4$. Results are displayed for $\Lambda=0.2$\hmpc, only.}
    	\label{tab:waicG3}
    	\begin{tabular}{lcccc}
    	   \hline
    	   Model & $\Gamma_3$ & $k^2$ & $\Delta$WAIC (NR \& RNL) & $\Delta$WAIC (RL) \\
    	   \hline
    	   \hline
           $\mathcal{M}_{5\mathrm{a}}$ & x & & 1.72 & 1.02\\ 
    	   \hline
           $\mathcal{M}_{5\mathrm{b}}$ & & x & 2.35 &1.88\\  
    	   \hline
           $\mathcal{M}_{6}$ & x & x & 1.66 & 1.41\\    	   
    	   \hline
    	\end{tabular}
\end{table}

A generic bias expansion includes all the possible operators allowed by symmetries. Here, we perform a Bayesian model comparison to investigate how many bias parameters are actually needed to fit $P_{\delta_\mathrm{h}\delta}(k)$ extracted from our simulations for $k<0.2$ \hmpc. We focus on the $M_3$ sample that shows the most prominent scale-dependent bias in Fig.~\ref{fig:powertransfer}. We start with fitting the simple linear-bias model $P_{\delta_\mathrm{h}\delta}(k) = b_1 P_{\delta\delta}(k)$. Then, we add several combinations of the terms $b_2 P_{\delta^2 \delta}(k)$, $b_{s^2} P_{s^2 \delta}(k)$ and $b_{\nabla^2 \delta} P_{\nabla^2 \delta\delta}(k)$ as summarized in Table~\ref{tab:waic}. Finally, motivated by the perturbative results presented in section~\ref{sec:model}, we also consider the terms $b_{\Gamma_3} P_{\Gamma_3 \delta}(k)$ and $\Upsilon_{2,\epsilon_1,\epsilon_{\mathrm m}}\, k^2$, as well as a combination of the two (see Table~\ref{tab:waicG3}).

We quantify the relative performance of each model by using the Widely Applicable Information Criterion \citep{watanabe2010asymptotic}, also known as the Watanabe-Akaike Information Criterion (WAIC). This method evaluates the `predictive accuracy' of a model, i.e. how useful the model will be in predicting new or future measurements. For finite and noisy data, this concept differs from the `goodness of fit' which quantifies how well a model describes the data that have been used to optimize the model parameters. The WAIC is a Bayesian method that generalizes the Akaike Information Criterion \citep[AIC,][]{akaike1973maximum} by averaging over the posterior distribution of the model parameters. It can be shown that it is asymptotically equivalent to Bayesian cross validation \citep{watanabe2010asymptotic}. Moreover, the WAIC can be seen as an improvement upon the Deviance Information Criterion \citep[DIC,][]{spiegelhalter2002bayesian} as it is invariant under re-parametrization of the model and also works for singular models (where the Fisher information matrix is not invertible). To apply the WAIC, one first estimates the log pointwise predictive density of the model \citep[lppd,][]{gelman2014understanding}
\begin{equation}
    \mathrm{lppd}=\sum_i \ln \langle P(\mathbf{w}_i|\boldsymbol{\theta})\rangle_\mathrm{post}
\end{equation}
where the sum runs over the data points $\mathbf{w}_i$, and $P(\mathbf{w}_i|\boldsymbol{\theta})$ denotes the probability to measure $\mathbf{w}_i$ under the statistical model and for a given set of model parameters $\boldsymbol{\theta}$. The expectation $\langle \dots \rangle_\mathrm{post}$ is estimated using draws for the model parameters  from the posterior distribution given by the MCMC chains.
The better the model fits the data, the larger the lppd is. In order to avoid overfitting, the lppd should be penalized for the effective number of model parameters, $p_\mathrm{WAIC}$.  Two different estimators can be used to measure this quantity from the MCMC chains, namely
\begin{equation}
 p_{\mathrm{WAIC},1}=2 \sum_i [\log \langle P(\mathbf{w}_i|\boldsymbol{\theta}) \rangle_\mathrm{post}-\langle \log P(\mathbf{w}_i|\boldsymbol{\theta})\rangle_\mathrm{post}]\;,   
 \label{eq:pwaic1}
\end{equation}
and
\begin{equation}
 p_{\mathrm{WAIC},2}=\sum_i 
 \langle [\log P(\mathbf{w}_i|\boldsymbol{\theta})-
 \langle \log P(\mathbf{w}_i|\boldsymbol{\theta})\rangle_\mathrm{post}
 ]^2
 \rangle_\mathrm{post} \;.
 \label{eq:pwaic2}
\end{equation}
Essentially, each model parameter counts as one if all the information about it comes from the likelihood function, as zero if all the information comes from the prior, or as an intermediate number whenever both the data and the prior are informative. For our fits, we do not notice any practical difference between using equation~(\ref{eq:pwaic1}) or equation~(\ref{eq:pwaic2}). Following \citet{gelman2014understanding}, we finally define the WAIC by using a deviance scale (i.e. by multiplying the lppd by a factor of 2 so that the final result is more easily comparable with the AIC and the DIC):
\begin{equation}
    \mathrm{WAIC}=-2 (\mathrm{lppd}-p_{\mathrm{WAIC},i})
    \label{eq:waic}
\end{equation}
The lower the WAIC is, the better the model performs.  It is generally understood that a difference $\Delta$WAIC of 5 (10) provides `suggestive' (`substantial') evidence in favour of the preferred model.

\subsection{Bare bias expansion}
\label{ssec:barebias}

When we do not apply any renormalisation (hereafter NR, short for `no renormalisation'), the lowest WAIC is obtained for model $\mathcal{M}_4$,
\begin{align}
P_{\delta_\mathrm{h}\delta}(k) = b_1 P_{\delta\delta}(k) & + b_{\nabla^2 \delta} P_{\nabla^2 \delta\delta}(k)\nonumber \\ &+ b_2 P_{\delta^2 \delta}(k)  + b_{s^2} P_{s^2 \delta}(k) \; ,
\label{eq:biasexpansion}
\end{align}
for all values of $\Lambda$.
In Table~\ref{tab:waic}, we report the difference $\Delta$WAIC between each model we have considered and the preferred one. Our results clearly rule out models with less than four bias parameters. On the other hand, considering additional terms proportional to $\Gamma_3$ and $k^2$ and their combination gives a WAIC which is slightly worse than for $\mathcal{M}_4$, as we present in Table~\ref{tab:waicG3} for $\Lambda=0.2$\hmpc (since higher-order corrections should be most important for this cutoff). This shows that there is no need to include $b_{\Gamma_3}$ and $b_{k^2}$ in the bias expansion as, for $k<0.2$\hmpc, our measurement of $P_{\delta_\mathrm{h}\delta}(k)$ within $\sim 70\,h^{-3}$ Gpc$^3$ cannot constrain them. We thus conclude that, to describe the $z=0$ cross spectral density of massive DM haloes and matter on these scales, we need to account for the linear bias $b_1$, the non-linear bias $b_2$, the tidal bias $b_{s^2}$ and the first higher-derivative bias $b_{\nabla^2 \delta}$.

In the top panel of Fig.~\ref{fig:contrib}, we plot the best-fitting model $\mathcal{M}_4$ (solid curve) and the corresponding residuals are shown in the narrow panel below (fourth from the top). The fit never deviates from the measurements in a statistically significant way and sub-percent accuracy is achieved for most values of $k$. In order to visually illustrate the need for and the concept of renormalisation, in the top panel, we also plot the individual contributions of the different terms appearing on the r.h.s. of
equation (\ref{eq:biasexpansion}). Note that $P_{\delta \delta}(k)$, $P_{\delta^2 \delta}(k)$ and $P_{s^2 \delta}(k)$ are proportional to each other when $k\to 0$. Moreover, the terms that scale with $b_2$ and
$b_{s^2}$ are never negligible compared with $b_1 P_{\delta\delta}(k)$, even at the largest scales we can probe.

\subsection{Renormalised bias expansion}
\label{ssec:renormbias}

\begin{figure}
  \centering{
      \includegraphics[height=0.7\textheight]{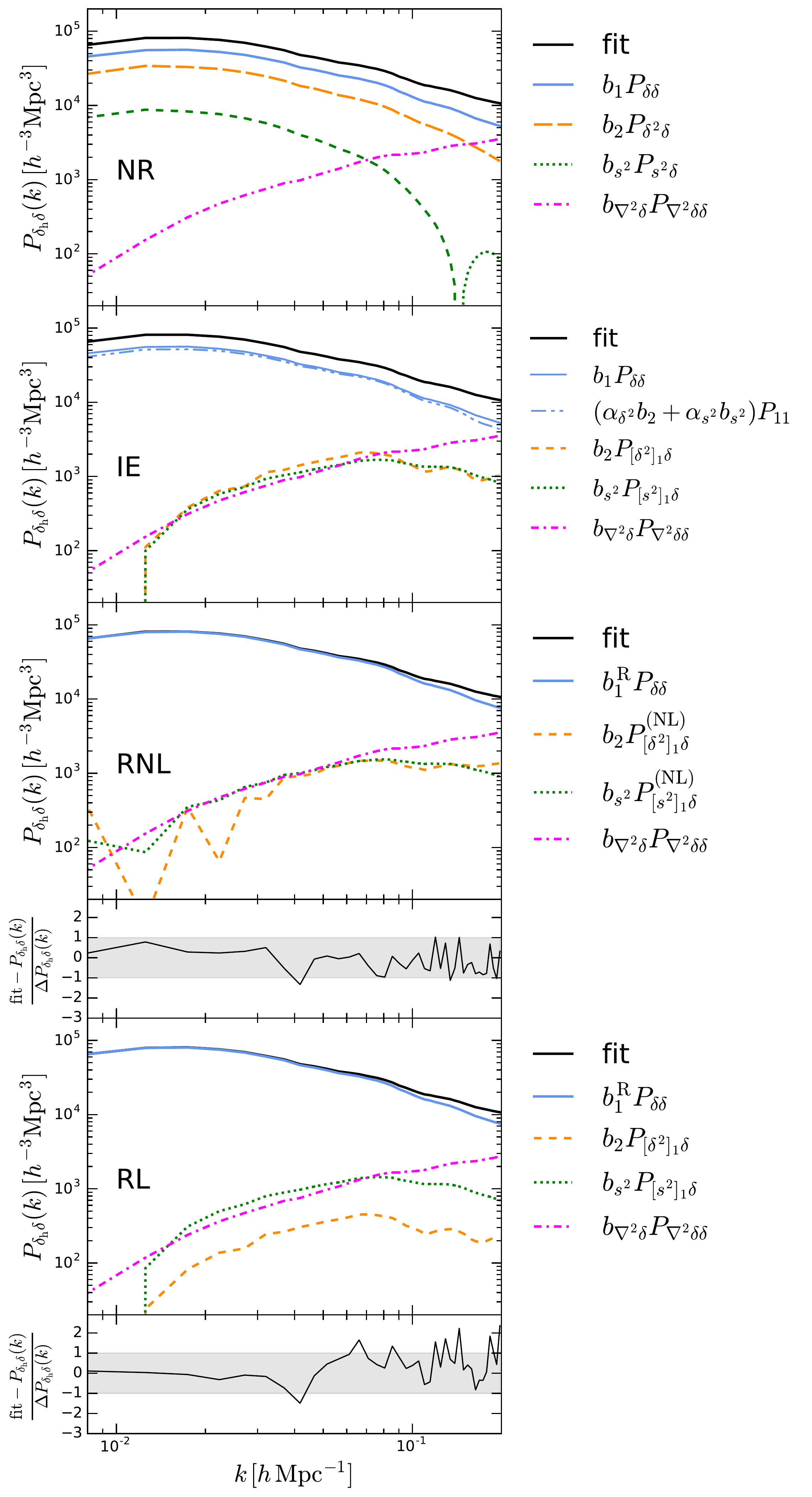}\\
    }
    \caption{The best fit of the bias model $\mathcal{M}_4$ to $P_{\delta_\mathrm{h}\delta}(k)$ (solid) and its different components (as indicated in the legend) for the halo mass bin $M_3$ and $\Lambda=0.2$\hmpc. The contribution proportional to $b_s^2$ in the top panel changes sign at $k\simeq 0.15$\hmpc and we plot its absolute value but use a dashed line to highlight where it is negative. Similarly, all the contributions proportional to $b_2$ in the bottom three panels are negative. To improve readability, we neglect the uncertainty in the bias parameters (only using their posterior mean). In the top three panels, we show the results obtained for the NR, IE and RNL cases which give rise to the same fit. The fit residuals
    normalised to the statistical error of the data $\Delta P_{\delta_\mathrm{h}\delta}(k)$ are displayed in the fourth panel.  Results for the RL case are displayed in the bottom two panels.}
    \label{fig:contrib}
\end{figure}

\begin{figure}
  \centering{
      \includegraphics[width=\columnwidth]{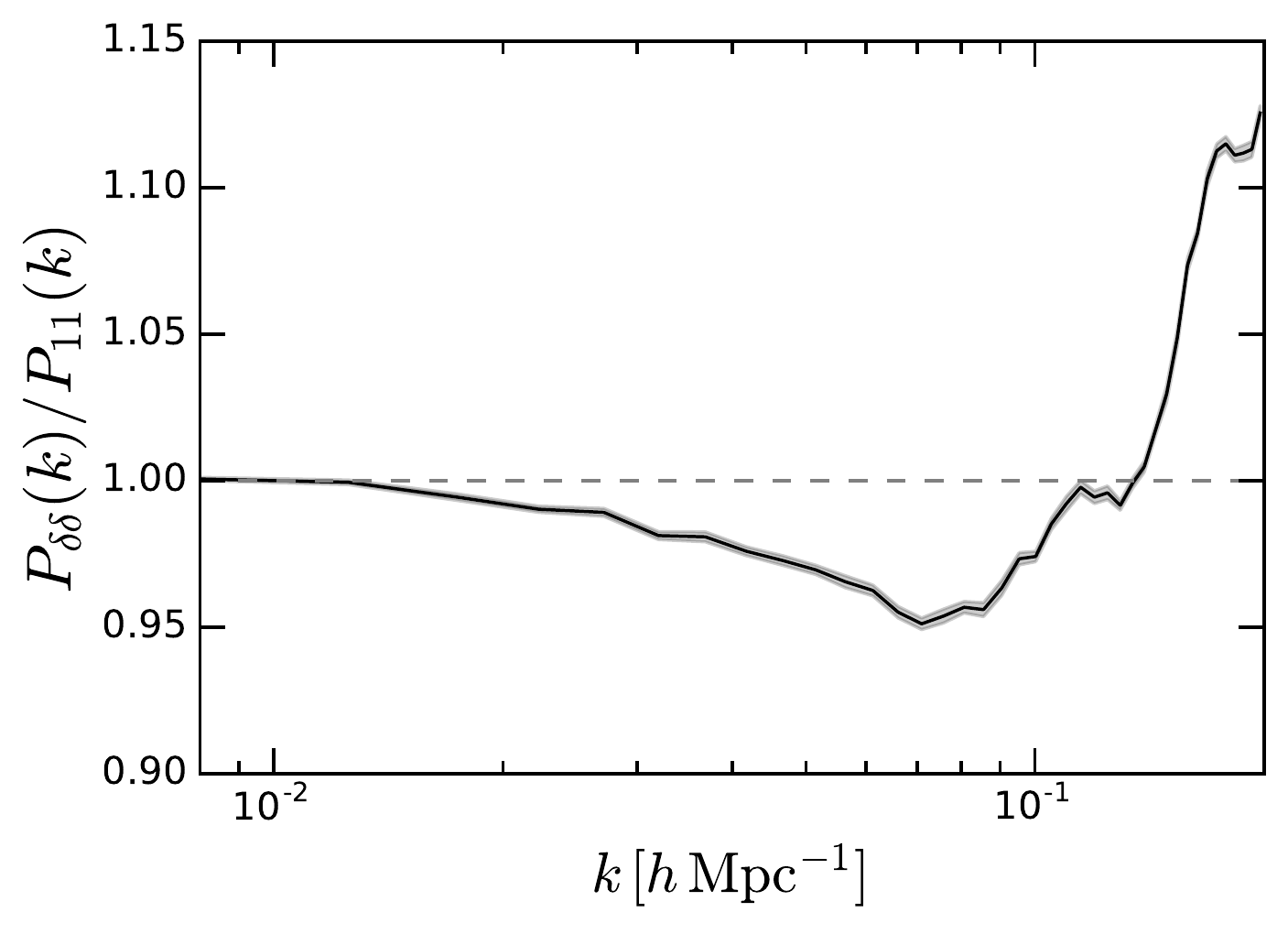}\\
    }
    \caption{Non-linear corrections to the matter power spectrum at $z=0$. The line shows the average ratio taken over the simulations. Shaded regions indicate the standard error of the mean.}
    \label{fig:nlvs11}
\end{figure}
  
For every model expressed in terms of the bare bias parameters, we can renormalise the cross-spectra of the composite operators with $\delta$ as in the examples provided by equations (\ref{eq:renormd2}) and (\ref{eq:renorms2}). 
The subtracted terms proportional to $P_{11}(k)$ will then contribute to the renormalised linear bias parameter. Let us consider, for example, equation (\ref{eq:biasexpansion}) and re-write it in terms of the renormalised cross spectra
\begin{align}
P_{\delta_\mathrm{h}\delta}(k) = b_1 P_{\delta\delta}(k) & + (\alpha_{\delta^2} b_2+\alpha_{s^2} b_{s^2})\, P_{11}(k)+ b_{\nabla^2 \delta} P_{\nabla^2 \delta\delta}(k) \nonumber \\ & +  b_2 P_{[\delta^2]_1 \delta}(k)+b_{s^2} P_{[s^2]_1 \delta}(k) \;,
\label{eq:biasexpansionrenshift}
\end{align}
with $\alpha_{\delta^2}$ and $\alpha_{s^2}$ the numerical coefficients
appearing in equations (\ref{eq:renormd2}) and (\ref{eq:renorms2}).
The result of this decomposition for the best-fitting model $\mathcal{M}_4$ is illustrated in the
second panel from the top of Fig.~\ref{fig:contrib}
(labelled IE, short for `intermediate expansion').
The renormalised cross spectra of the composite operators are now subdominant (and have a different
shape) for $k\to 0$ with respect to the linear bias terms. Their relative contributions to $P_{\delta_\mathrm{h}\delta}(k)$ grow with $k$ (although their sum nearly vanishes in this particular case).

In perturbative calculations, the spectrum $P_{11}(k)$ appearing in equation~(\ref{eq:biasexpansionrenshift}) is promoted to $P_{\delta \delta}(k)$ to write
\begin{align}
P_{\delta_\mathrm{h}\delta}(k) = b^\mathrm{R}_1 P_{\delta\delta}(k)  &+ b_{\nabla^2 \delta} P_{\nabla^2 \delta\delta}(k)  \nonumber \\ & + b_2 P_{[\delta^2]_1 \delta}(k)+ b_{s^2} P_{[s^2]_1 \delta}(k)\; ,
\label{eq:biasexpansionRL}
\end{align}
which is the analog of equation (\ref{eq:fullpert}) with
\begin{equation}
b_1^\mathrm{R}=b_1+\alpha_{\delta^2} b_2+\alpha_{s^2} b_{s^2}\, . 
\label{eq:b1theo4p}
\end{equation}
In the simulations, however, this expression does not coincide any longer with equation~(\ref{eq:biasexpansionRL}) as $P_{11}(k)\neq P_{\delta \delta}(k)$ for $k>0$ (see Fig.~\ref{fig:nlvs11}). 
The only option to renormalise the linear bias coefficient without altering equation~(\ref{eq:biasexpansion}) is to generalize equation (\ref{eq:renormd2}) to the non-linear regime by using 
\begin{align}
P_{[\delta^2]_1 \delta}^{(\mathrm{NL})}(k) = P_{\delta^2 \delta}(k) - \alpha_{\delta^2}(\Lambda)\,P_{\delta\delta}(k)\;,
\label{eq:renNL}
\end{align}
together with analogous relations for the other composite operators. We can thus re-write equation~(\ref{eq:biasexpansion}) as 
\begin{align}
P_{\delta_\mathrm{h}\delta}(k) = b^\mathrm{R}_1 P_{\delta\delta}(k)  &+ b_{\nabla^2 \delta} P_{\nabla^2 \delta\delta}(k)  \nonumber \\ & + b_2 P_{[\delta^2]_1 \delta}^{(\mathrm{NL})}(k)+ b_{s^2} P_{[s^2]_1 \delta}^{(\mathrm{NL})}(k)
\label{eq:biasexpansionRNL}
\end{align}
which we refer to as the RNL, short for `non-linear renormalisation'.
In the third panel from the top of Fig.~\ref{fig:contrib},
we show the best-fitting model $\mathcal{M}_4$ and its different components. Obviously, fitting the RNL expression to the numerical data is completely
equivalent to fitting the NR one and gives exactly the same constraints on the non-linear and higher-derivative bias parameters. In fact, all the spectra that are used in the fits are the same,
it is only their labelling that changes. In the NR case, the terms $\alpha_{\delta^2}P_{\delta\delta}$ and $\alpha_{s^2}P_{\delta\delta}$ are considered as parts of $P_{\delta^2\delta}(k)$ and $P_{s^2\delta}(k)$, respectively, whereas they are interpreted as part of the renormalised linear bias term in the RNL case. Therefore, the conclusions regarding the optimal number of bias parameters that we have drawn for the bare bias expansion still apply to the RNL case.

On the other hand, in hands-on situations, it is common practice to fit one-loop perturbative models to survey or simulation data \citep[e.g.][]{saito2014understanding}. In our study, taking model $\mathcal{M}_4$ as an example, this corresponds to using equation (\ref{eq:biasexpansionRL}) together with (\ref{eq:renormd2}) and (\ref{eq:renorms2}). In this case (hereafter dubbed RL, short for `linear renormalisation'), the best-fitting parameters and the goodness of fit do not coincide with those obtained for NR and RNL models.
We thus repeat our model-selection test focusing on RL models. We find that the preferred operator sets vary with the cutoff scale (see the rightmost three columns in Table~\ref{tab:waic}).
For $\Lambda=0.05$ and 0.1 \hmpc, all models that contain a term in $\nabla^2\delta$ are favoured. For  $\Lambda=0.2$\hmpc, instead, $\mathcal{M}_{4}$ is singled out by the WAIC (the bottom two panels of Fig.~\ref{fig:contrib} display the individual components of $\mathcal{M}_{4}$ and the fit residuals).  
Further adding the $\Gamma_3$ and/or $k^2$ operators leads to slightly larger WAIC indicating mild overfitting. These results suggest that, as expected in an effective theory, more bias parameters are needed if smaller scales are considered. In particular, while $\delta$ and $\nabla^2 \delta$ form the minimal set of operators required for $\Lambda<0.1$\hmpc,  adding $[\delta^2]_1$ and $[s^2]_1$ is necessary to describe the data with $\Lambda=0.2$\hmpc.

\subsection{Bias parameters}
\label{ssec:biasfit}
  
\begin{table*}
    	\centering
    	\caption{The bias parameters obtained by fitting equations~(\ref{eq:biasexpansion}) and (\ref{eq:biasexpansionRNL}) to $P_{\delta_\mathrm{h}\delta}(k)$ from our 40 simulations.}
    	\label{tab:biasparamNR}
    	\begin{tabular}{ccccccc}
    	   \hline
    	   Mass & $\Lambda$ & $b_1$ & $b_1^\mathrm{R}$ & $b_2$ & $b_{s^2}$ & $b_{\nabla^2 \delta}$ \\
    	   bin & $[h\, \mathrm{Mpc}^{-1}]$ & (NR) & (RNL) & & & $[h^2\, \mathrm{Mpc}^{-2}]$ \\
    	   \hline
    	   \hline
    	   & $0.05$ & $1.91 \pm 0.02$ & $1.89 \pm 0.01$ & $-0.09 \pm 0.26$ & $-0.14 \pm 0.46$ & $5.53 \pm 7.34$ \\
    	   $M_1$ & $0.1$ & $1.91 \pm 0.02$ & $1.89 \pm 0.01$ & $-0.003 \pm 0.10$ & $-0.08 \pm 0.08$ & $7.12 \pm 1.26$ \\
    	   & $0.2$ & $2.04 \pm 0.04$ & $1.895 \pm 0.003$ & $-0.05 \pm 0.03$ & $-0.09 \pm 0.01$ & $2.45 \pm 0.36$ \\   
    	   \hline
    	   & $0.05$ & $2.38 \pm 0.02$ & $2.39 \pm 0.02$ & $0.14 \pm 0.39$ & $0.06 \pm 0.67$ & $17.87 \pm 10.28$ \\
    	   $M_2$ & $0.1$ & $2.30 \pm 0.04$ & $2.38 \pm 0.01$ & $0.38 \pm 0.15$ & $-0.25 \pm 0.12$ & $18.34 \pm 1.82$ \\ 
    	   & $0.2$ & $2.35 \pm 0.05$ & $2.378 \pm 0.004$ & $0.12 \pm 0.05$ & $-0.15 \pm 0.01$ & $10.71 \pm 0.55$ \\    
    	   \hline
    	   & $0.05$ & $3.26 \pm 0.03$ & $3.35 \pm 0.02$ & $2.92 \pm 0.55$ & $-2.79 \pm 0.95$ & $57.75 \pm 15.00$ \\
    	   $M_3$ & $0.1$ & $2.94 \pm 0.05$ & $3.36 \pm 0.01$ & $1.87 \pm 0.20$ & $-1.00 \pm 0.15$ & $54.50 \pm 2.39$ \\
    	   & $0.2$ & $2.35 \pm 0.07$ & $3.35 \pm 0.01$ & $1.07 \pm 0.06$ & $-0.42 \pm 0.01$ & $40.33 \pm 0.67$ \\
    	   \hline
    	\end{tabular}
\end{table*}

\begin{table*}
    	\centering
    	\caption{As in Table~\ref{tab:biasparamNR} but using equation~(\ref{eq:biasexpansionRL}).
    	}
    	\label{tab:biasparamR1L}
    	\begin{tabular}{cccccc}
    	   \hline
    	   Mass & $\Lambda$ & $b_1^\mathrm{R}$ & $b_2$ & $b_{s^2}$ & $b_{\nabla^2 \delta}$ \\
    	   bin & $[h\, \mathrm{Mpc}^{-1}]$ &  &  &  & $[h^2\, \mathrm{Mpc}^{-2}]$ \\
    	   \hline
    	   \hline
    	   & $0.05$ & $1.90 \pm 0.01$ & $-0.10 \pm 0.27$ & $0.01 \pm 0.05$ & $7.81 \pm 6.09$ \\
    	   $M_1$ & $0.1$ & $1.90 \pm 0.01$ & $-0.10 \pm 0.08$ & $0.02 \pm 0.08$ & $7.34 \pm 0.96$ \\
    	   & $0.2$ & $1.898 \pm 0.002$ & $0.05 \pm 0.01$ & $-0.10 \pm 0.01$ & $3.84 \pm 0.13$ \\
    	   \hline
    	   & $0.05$ & $2.38 \pm 0.01$ & $-0.14 \pm 0.40$ & $-0.05 \pm 0.07$ & $12.03 \pm 8.58$ \\
    	   $M_2$ & $0.1$ & $2.38 \pm 0.01$ & $0.18 \pm 0.12$ & $-0.14 \pm 0.13$ & $17.33 \pm 1.39$ \\
    	   & $0.2$ & $2.379 \pm 0.004$ & $0.12 \pm 0.01$ & $-0.15 \pm 0.01$ & $10.56 \pm 0.17$ \\
    	   \hline
    	   & $0.05$ & $3.35 \pm 0.02$ & $-0.87 \pm 0.62$ & $-0.07 \pm 0.11$ & $42.85 \pm 13.53$ \\
    	   $M_3$ & $0.1$ & $3.36 \pm 0.01$ & $-0.05 \pm 0.17$ & $-0.07 \pm 0.17$ & $41.46 \pm 1.96$ \\
    	   & $0.2$ & $3.35 \pm 0.01$ & $0.23 \pm 0.02$ & $-0.36 \pm 0.02$ & $30.97 \pm 0.23$ \\
    	   \hline
    	\end{tabular}
\end{table*}
  
The best-fitting values of the bias parameters obtained for the NR and the RNL cases are displayed in Table~\ref{tab:biasparamNR} for all mass bins and cutoff scales. We show both the bare linear bias parameter ($b_1$) and the renormalised one ($b_1^\mathrm{R}$). Similarly, in Table~\ref{tab:biasparamR1L}, we report the best-fitting parameters for the RL fits. Notice that, while $b_1$ is $\Lambda$-dependent, $b_1^\mathrm{R}$ assumes consistent values for all cutoffs in both the RNL and RL cases. We can thus conclude that the non-perturbative renormalisation procedures we have implemented in the simulations were completely successful. Nevertheless, it is worth pointing out that the best-fitting values of
$b_2$, $b_{s^2}$ and $b_{\nabla^2 \delta}$ in Tables~\ref{tab:biasparamNR} and \ref{tab:biasparamR1L} differ for all mass bins when $\Lambda \geq 0.1\,h\, \mathrm{Mpc}^{-1}$. In fact, the difference between $P_{11}(k)$ and $P_{\delta\delta}(k)$ becomes substantial at small
scales where also the signal generated by the non-linear and derivative bias terms is important. Therefore, the RL fit compensates for the improper renormalisation of the linear-bias term by spuriously altering $b_2$, $b_{s^2}$ and $b_{\nabla^2 \delta}$. These results demonstrate the limitations of applying perturbative renormalisation at finite values of $k$.
As already discussed in section~\ref{ssec:renormbias}, in order to retain the measurements of $b_2$, $b_{s^2}$ and $b_{\nabla^2 \delta}$ from the NR fit in the renormalised case, equation~(\ref{eq:renNL}) (and its analogue for the other operators) must be used instead of (\ref{eq:renormd2}).

\subsection{Comparison with previous work}
\label{sec:othermodels}

\begin{figure*}
  \centering{
      \includegraphics[width=0.94\textwidth]{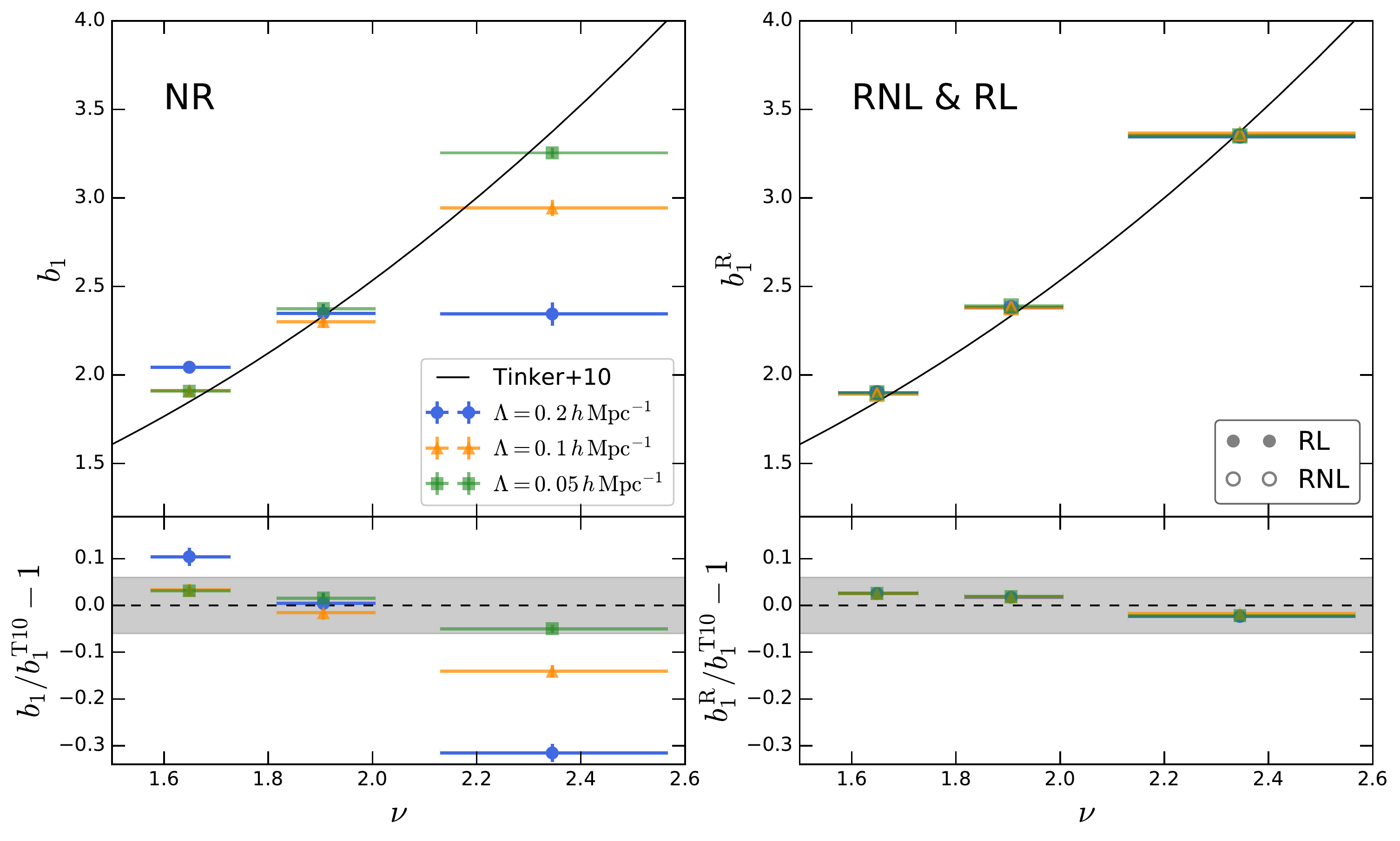}\\
    }
    \caption{The best-fitting linear bias parameter as a function of the peak height, $\nu$, for different cutoff scales, $\Lambda$. The left-hand side panel refers to the NR case in which $b_1$ runs with $\Lambda$. On the contrary, the right-hand side panel shows the results obtained by fitting $b_1^\mathrm{R}$. Although solid and open symbols characterize the RNL and RL cases, respectively, all symbols referring to the same mass bin overlap and cannot be distinguished. As a reference, we also show the fitting function by \citet{tinker2010large} (solid) and its relative deviation from our results (bottom panels). The shaded area represents the intrinsic scatter about the mean relation found by \citet{tinker2010large}.}
    \label{fig:results-b1}
\end{figure*}
  
\begin{figure*}
  \centering{
      \includegraphics[width=0.94\textwidth]{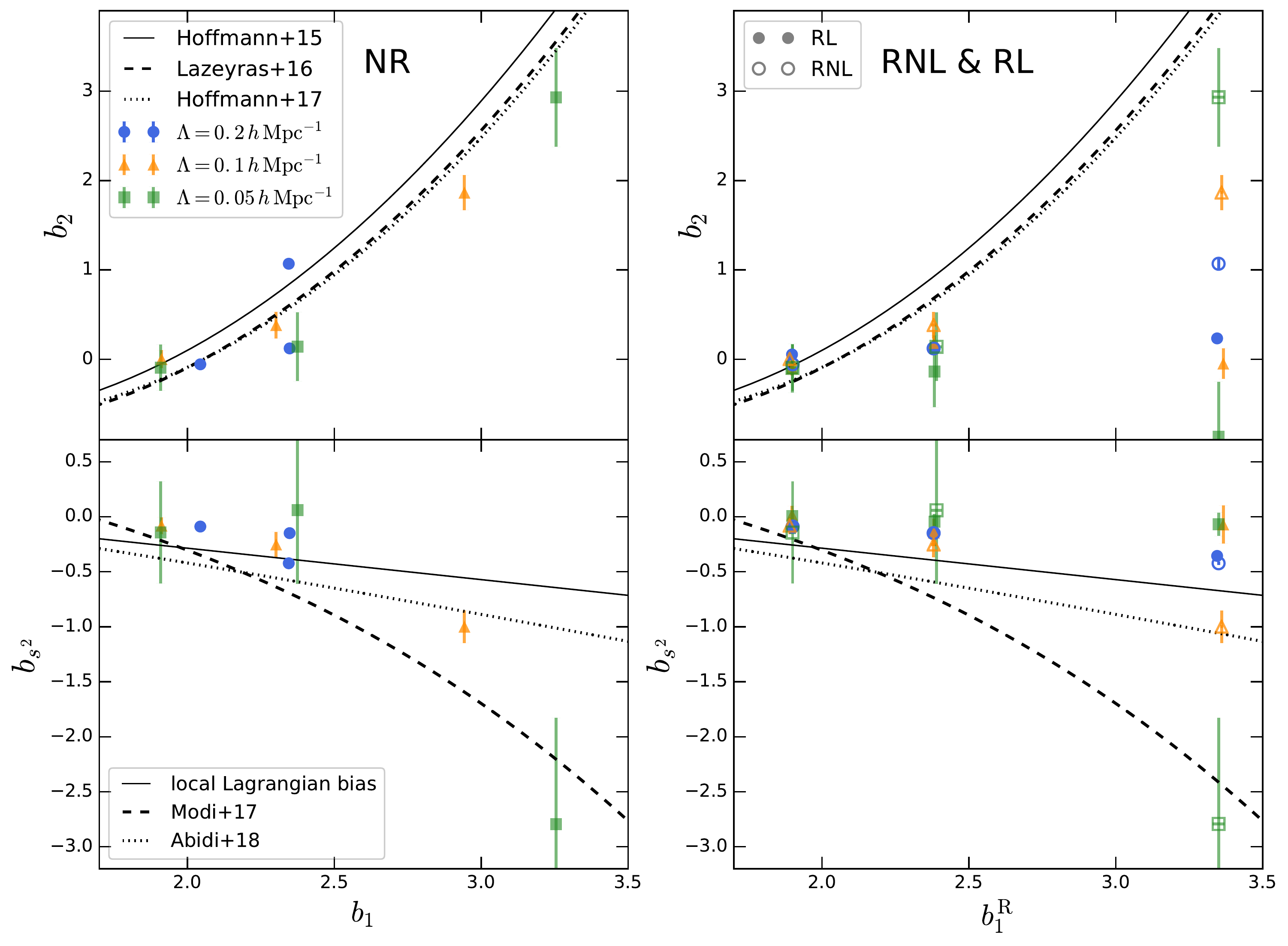}\\
    }
    \caption{Symbols with errorbars indicate the best-fitting non-linear bias parameters $b_2$ (top) and $b_{s^2}$ (bottom) as a function of $b_1$ for different cutoff scales. The panels on the left-hand side show the bare bias parameters (NR) while the results on the right-hand side are obtained after renormalising $b_1$ (here, empty and filled symbols correspond the RNL and RL cases, respectively). As a reference, we overplot several theoretical models and fitting functions for the renormalised parameters (see the main text for details).}
    \label{fig:results-b2bs2}
\end{figure*}

\begin{figure*}
  \centering{
      \includegraphics[width=0.94\textwidth]{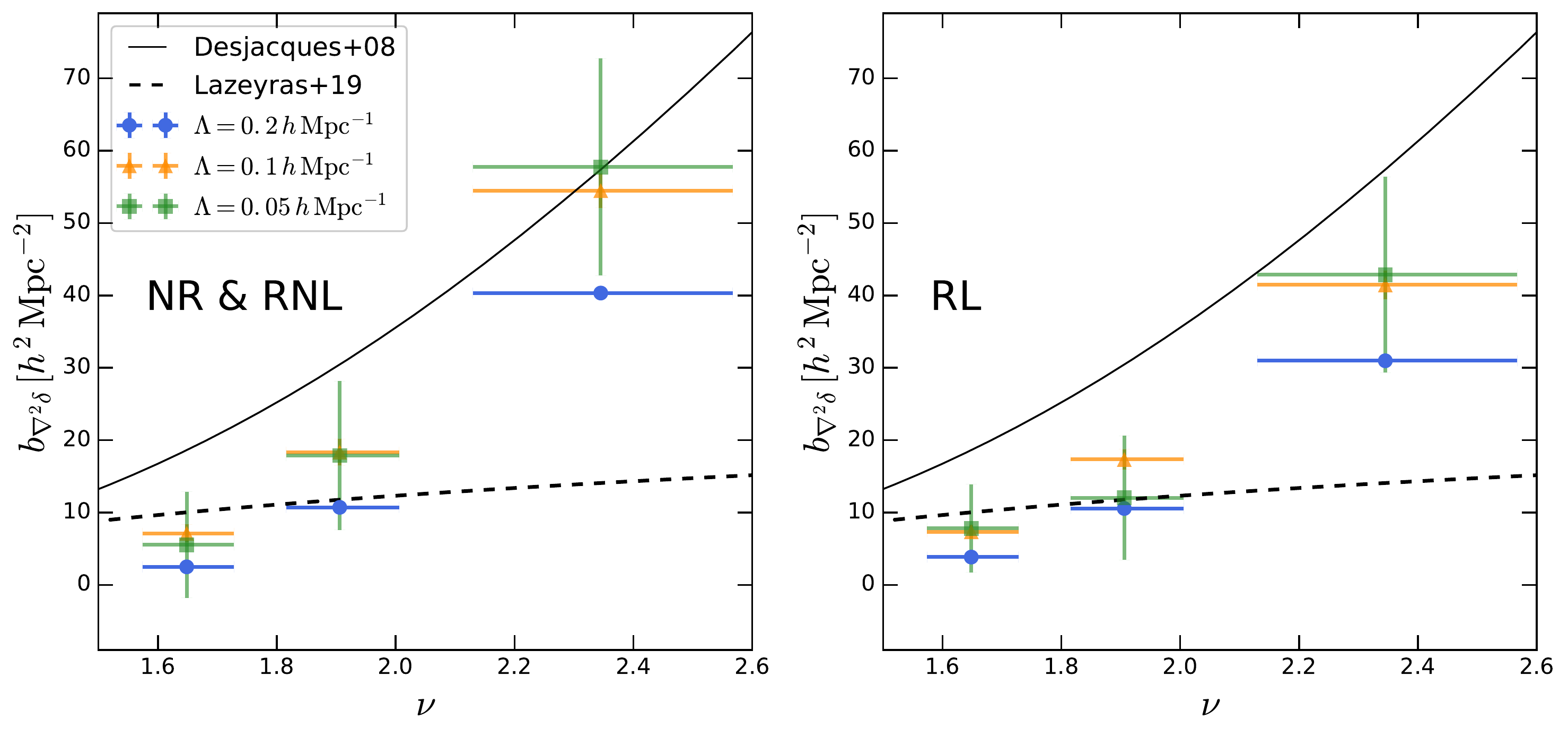}\\
    }
    \caption{The best-fitting first higher-derivative bias parameter, $b_{\nabla^2 \delta}$, as a function of the peak height, $\nu$, for different cutoff scales.
    As a reference, we show the the fit to numerical simulations by \citet[][dashed]{lazeyras2019robust} and the 
    predictions from peak statistics presented in \citet[][solid]{desjacques2008baryon}.}
    \label{fig:results-bnabla2}
\end{figure*}

We compare the halo bias parameters we obtained from our simulations with previous results in the literature. In Fig.~\ref{fig:results-b1}, we contrast our results for $b_1$ and $b_1^\mathrm{R}$ (symbols) with the fit by \cite{tinker2010large} -- their equation (6) -- evaluated for a mean halo overdensity of $\Delta=333$ (solid line). The fit quantifies the relative amplitude of the halo and matter power spectra on large scales as measured from a set of N-body simulations with slightly different cosmological parameters. To ease the comparison, we express the mass dependence in terms of the `peak height' $\nu=\delta_\mathrm{c}/\sigma_M$. Here, $\sigma_M$ denotes the linear rms density fluctuation smoothed over the Lagrangian patch of each halo using a spherical top-hat filter and $\delta_\mathrm{c}=1.686$ is the critical linear overdensity in the spherical collapse model (for an Einstein-de Sitter universe). The horizontal location of the symbols is obtained by computing $\nu$ for every halo and averaging over each mass bin. In the NR case (left), our best-fitting values run with $\Lambda$ and differ from the fit by \citet{tinker2010large} by up to 30 per cent. After renormalisation (right), the $\Lambda$-independent data points obtained in the RNL and RL cases are in  excellent agreement with each other and also match the results by \citet{tinker2010large} to better than 2 per cent which is a factor of three smaller than the intrinsic scatter about the mean relation between linear bias and halo mass found in \citet{tinker2010large}. This shows that by renormalising the spectra we automatically obtain a consistent and robust value of $b_1^\mathrm{R}$ for all the mass bins.

In Fig.~\ref{fig:results-b2bs2}, we compare our measurements of $b_2$ and $b_{s^2}$  with other results in the literature. For the quadratic bias (top panels), we consider three different fitting functions for the relation $b_2(b_1)$ extracted from N-body simulations.
The first was obtained by applying the peak-background split to measurement of the halo mass function \citep{hoffmann2015comparing}. The second was determined with the `separate-universe' technique  
in which one studies the response to infinite-wavelength perturbations \citep{lazeyras2016precision}. 
The third, instead, was derived from measurements of the halo reduced three-point correlation function
\citep{hoffmann2017linear}. 
Note that these methods measure the renormalised $b_2$ while we only renormalise $b_1$. Therefore, we do not expect that our $\Lambda$-dependent best-fitting values for $b_2$ should perfectly coincide with the other results in the literature. Nevertheless, in the NR case (left panel) and for large-scale cutoffs (i.e $\Lambda=0.05$ and 0.1\hmpc), our measurements approximately run along the fitting functions for the relation $b_2(b_1)$.
After renormalising $b_1$ (right panel), however,
our $b_2$ measurements for the two highest mass bins tend to lie below the previously published results. Moreover, as already mentioned, we obtain different $b_2$ values in the RNL and RL cases. The RNL results with small $\Lambda$ are in much better agreement with the fitting functions in the literature. This suggests that it should be possible to further improve the agreement by extending the RNL technique to renormalise $b_2$. We will explore this possibility in our future work.
For the tidal bias (bottom panels), we also compare our best-fitting values to three different literature results. We consider the theoretical prediction for a local Lagrangian biasing scheme \citep{catelan2000two}, $b_{s^2}(b_1)=-(2/7)\, (b_1-1)$ \citep{baldauf2012evidence, chan2012gravity} together with two fitting functions that account for a non-vanishing Lagrangian tidal-tensor bias in N-body simulations \citep[][their equations 22 and 5.1 plus 2.32, respectively]{modi2017halo,abidi2018cubic}.
In agreement with these fitting functions, our measurements provide evidence for large negative
tidal biases at high halo masses. After renormalising $b_1$, this conclusion holds only in the RNL case while $b_{s^2}$ always assumes values close to zero in the RL case.

In Fig.~\ref{fig:results-bnabla2}, we plot our results for $b_{\nabla^2 \delta}$ as a function of $\nu$. The NR and RNL fits are shown together in the left-hand side panel since $b_{\nabla^2 \delta}$ assumes the same value in these cases. On the contrary, the RL results are plotted in the right-hand side panel.
As a benchmark, we also show the fitting function obtained with the `amplified-mode' simulation technique by \citet[][their equation 5.4 but note that we adopt the opposite sign convention for $b_{\nabla^2 \delta}$]{lazeyras2019robust} as well as
the predictions for density peaks \citep{desjacques2008baryon} computed using the filter function given in \citet{chan2015effective}.
In agreement with \citet{elia2012spatial}, our measurements for the haloes scale with $\nu$ in a similar way as the theoretical peak model but assume lower values in the $\nu$-range we can probe. Compared with the fit by \citet{lazeyras2019robust}, we find comparable results in bins $M_1$ and $M_2$ and substantially higher values for $M_3$.  Overall our NR and RNL measurements display a steeper $\nu$-dependence than expected based on their results.

\section{Summary}
\label{sec:concl}
We have investigated a number of issues related to the clustering of biased tracers of the LSS. In particular, we have focused on the renormalisation of the linear bias parameter. After reviewing the literature on the subject, we have applied a bias expansion to N-body simulations and studied the UV-sensitivity of the composite operators that contribute to the halo-matter cross spectrum $P_{\delta_\mathrm{h}\delta}(k)$. We have then successfully mastered the challenge to renormalize these terms without resorting to perturbation theory. Finally, we have identified how many bias parameters are needed to accurately describe halo clustering for $k<0.2$\hmpc without overfitting.
Our main results can be summarized as follows.
\begin{enumerate}
    \item We have run a suite of 40 N-body simulations and measured $P_{\delta_\mathrm{h}\delta}(k)$ for three different halo-mass bins selected at $z=0$. Consistently with previous work, we have found that the ratio $P_{\delta_\mathrm{h}\delta}(k)/P_{\delta\delta}(k)$  grows with $k$ and that this effect becomes more prominent for more massive haloes (Fig.~\ref{fig:powertransfer}). The scale dependence of the ratio provides a compelling motivation for considering non-linear bias models.
    \item We have measured all fields that enter $P_{\delta_\mathrm{h}\delta}(k)$ at one loop in PT -- equation~(\ref{eq:fullpert}) -- using different values of the coarse-graining scale ($\Lambda=0.05, 0.1$ and 0.2\hmpc, Fig.~\ref{fig:simlambda}). By cross-correlating these fields with the mass overdensity, we have computed all partial contributions to $P_{\delta_\mathrm{h}\delta}(k)$. Focusing on the composite operators appearing in the bias expansion (e.g. $\delta^2$ and $s^2$), we have shown that the amplitude of their cross spectra with $\delta$ strongly depends on $\Lambda$  (top panels of Fig.~\ref{fig:spectraall}). We have demonstrated that this `UV-sensitivity' is fully captured by the cross correlation between the composite operators and the linear density field (middle panels of Fig.~\ref{fig:spectraall}). Thus, we have successfully obtained the renormalised spectra by subtracting this UV-sensitive term from the original spectra (bottom panels of Fig.~\ref{fig:spectraall}).
    \item We have compared $P_{\delta^2 \delta_1}(k)$ and $P_{s^2 \delta_1}(k)$ extracted from the simulations to perturbative predictions at LO. An excellent agreement over a broad range of scales can be achieved by properly accounting for the window functions that define the coarse-graining procedure in the perturbative integrals (top panels of Fig.~\ref{fig:renormtest}). The values $68/21\,\sigma_1^2$ and $136/63\,\sigma_1^2$, that are usually quoted for the limit $k\to 0$, overpredict our measurements, as they are obtained neglecting the influence of the window function. The relative difference, however, decreases with increasing $\Lambda$.
    \item For $\Lambda=0.05$ and 0.1 \hmpc, we have shown that the `renormalised spectra' $P_{[\delta^2]_1 \delta}(k)$ and $P_{[s^2]_1 \delta}(k)$ measured in the simulations are in very good agreement with the perturbative calculations obtained by taking the window functions into account (bottom panels of Fig.~\ref{fig:renormtest}). For $\Lambda=0.2$ \hmpc, however, the 1-loop results show non-negligible deviations from the numerical data.
    \item We have fit bias models with a different number of parameters to $P_{\delta_\mathrm{h}\delta}(k)$ by using a Bayesian method that accounts for correlations between all spectra entering the model. With a model-selection criterion (WAIC), we have determined the optimal number of bias parameters that are needed as a function of $\Lambda$ for both the renormalised and non-renormalised spectra. The preferred set of bias operators includes $\delta, \nabla^2 \delta, \delta^2$ and $s^2$ in most cases (model $\mathcal{M}_4$ in Table~\ref{tab:waic}).
    \item In Fig.~\ref{fig:contrib}, we have illustrated the renormalisation procedure of the linear bias parameter for the optimal model. We have identified two different modi operandi. The first (RL) parallels perturbative renormalisation while the second (RNL) generalizes the renormalisation procedure to the fully non-linear regime.
    \item  We have presented the fits we obtained for the optimal bias model (see Tables~\ref{tab:biasparamNR} and \ref{tab:biasparamR1L}, as well as Figs.~\ref{fig:results-b1} -- \ref{fig:results-bnabla2}) and compared them to previous results in the literature. While the best-fitting values for the bare linear bias $b_1$ run with $\Lambda$, those for the renormalised linear bias $b_1^\mathrm{R}$ stay constant as a function of the cutoff scale (for both RL and RNL). This confirms that our numerical renormalisation was successful.
    \item Finally, we have shown that only RNL leaves the non-linear and higher-derivative bias parameters unchanged with respect to the bare bias expansion (NR). This casts some doubts on the accuracy and robustness of using NLO perturbative expressions to fit low-redshift observational and N-body data at $k\gtrsim 0.1$ \hmpc.
\end{enumerate}

\section*{Acknowledgements}

KW acknowledges partial support from the Transregional Collaborative Research Centre TRR 33 - The Dark Universe of the Deutsche Forschungsgesellschaft (DFG).


\bibliographystyle{mnras}
\bibliography{bibo}


\bsp	
\label{lastpage}
\end{document}